\IfFileExists{revtex4-2.cls}{%
  \documentclass[prl,twocolumn,superscriptaddress,showpacs]{revtex4-2}\def\REVTEX{}}{%
  \documentclass[twocolumn,10pt]{article}\usepackage[margin=0.75in]{geometry}}
\ifdefined\REVTEX\else
  \providecommand{\affiliation}[1]{\\{\small\itshape #1}}\providecommand{\email}[1]{}
  \newenvironment{acknowledgments}{\paragraph*{Acknowledgments.}}{}
\fi
\usepackage{graphicx,amsmath,amssymb,xcolor,booktabs,float,adjustbox}
\providecommand{\onecolumngrid}{\onecolumn}
\usepackage[colorlinks=true,allcolors=blue]{hyperref}

\providecommand{\semion}[1]{{\color{blue}[#1]}}
\newcommand{\ve}{\nu_e}
\ifdefined\REVTEX\else\renewcommand{\affiliation}[1]{\date{\small\itshape #1}}\fi
\begin{document}
\ifdefined\REVTEX
\title{Stabilized negative viscosity governs ocean jets before, during and after they leave the coast}
\author{Boris Galperin}
\email{bgalperin@usf.edu}
\affiliation{College of Marine Science, University of South Florida, St.\ Petersburg, Florida 33701, USA}
\author{Alexei Chekhlov}
\author{Ilya Staroselsky}
\affiliation{DS Americas Corp., Waltham, Massachusetts 02451, USA}
\author{Semion Sukoriansky}
\affiliation{Department of Mechanical Engineering, Ben-Gurion University of the Negev, Beer-Sheva 8410501, Israel}
\begin{abstract}
Eddies in rotating turbulence can feed the jets that spawn them, resulting in the negative viscosity
phenomenon. Taken alone, a negative viscosity represents up-gradient eddy fluxes and is often referred
to as ``antifriction'' or ``antidiffusion''. Mathematically, it leads to an ill-posed problem. In 
two-dimensional turbulence theory, this problem can be resolved by representing the transfer from small
to large scales by a \emph{stabilized} negative viscosity (SNV), in which a negative Laplacian viscosity
is balanced by its positive biharmonic counterpart. In purely two-dimensional turbulence, by the Taylor identity, a negative
viscosity is up-gradient transport of vorticity, which mixing cannot produce; on a slope, mixing of potential vorticity can mimic it, and we test for this. Using two years of wide-swath satellite altimetry (SWOT), we measure the eddy 
momentum flux across the five subtropical western boundary currents in each current's own frame. Along the 
continental slope the eddies act with a negative viscosity of order $-10^{3}$~m$^{2}$\,s$^{-1}$, with the 
SNV structure resolved in the Gulf Stream and the Agulhas. At separation from the slope the sign reverses 
in all five currents, and the eddies brake the flow. Downstream, in the free jets, the negative viscosity 
returns on the cyclonic flank, strengthened to $-10^{4}$~m$^{2}$\,s$^{-1}$. Separation is where the current 
leaves the strong topographic $\beta$ of the slope, the constraint that orders the eddy transfer. In an 
eddy-resolving reanalysis the SNV form holds across three ocean basins, with its shape parameter set 
by the deformation radius.

\end{abstract}
\maketitle
\else
\title{Stabilized negative viscosity governs ocean jets before, during and after they leave the coast}
\author{Boris Galperin$^{1}$ and Alexei Chekhlov$^{2}$ and Ilya Staroselsky$^{2}$ and Semion Sukoriansky$^{3}$}
\date{\small\itshape $^1$College of Marine Science, University of South Florida, St.\ Petersburg, Florida 33701, USA\\
$^2$DS Americas Corp., Waltham, Massachusetts 02451, USA\\
$^3$Department of Mechanical Engineering, Ben-Gurion University of the Negev, Beer-Sheva 8410501, Israel\\[4pt]
\upshape Draft 4, \today}
\maketitle
\begin{abstract}Eddies in rotating turbulence can feed the jets that spawn them, resulting in the negative viscosity
phenomenon. Taken alone, a negative viscosity represents up-gradient eddy fluxes and is often referred
to as ``antifriction'' or ``antidiffusion''. Mathematically, it leads to an ill-posed problem. In 
two-dimensional turbulence theory, this problem can be resolved by representing the transfer from small
to large scales by a \emph{stabilized} negative viscosity (SNV), in which a negative Laplacian viscosity
is balanced by its positive biharmonic counterpart. In purely two-dimensional turbulence, by the Taylor identity, a negative
viscosity is up-gradient transport of vorticity, which mixing cannot produce; on a slope, mixing of potential vorticity can mimic it, and we test for this. Using two years of wide-swath satellite altimetry (SWOT), we measure the eddy 
momentum flux across the five subtropical western boundary currents in each current's own frame. Along the 
continental slope the eddies act with a negative viscosity of order $-10^{3}$~m$^{2}$\,s$^{-1}$, with the 
SNV structure resolved in the Gulf Stream and the Agulhas. At separation from the slope the sign reverses 
in all five currents, and the eddies brake the flow. Downstream, in the free jets, the negative viscosity 
returns on the cyclonic flank, strengthened to $-10^{4}$~m$^{2}$\,s$^{-1}$. Separation is where the current 
leaves the strong topographic $\beta$ of the slope, the constraint that orders the eddy transfer. In an 
eddy-resolving reanalysis the SNV form holds across three ocean basins, with its shape parameter set 
by the deformation radius.
\end{abstract}
\fi

\textit{Introduction.}---In 1786 Benjamin Franklin advised captains sailing from Newfoundland to New York to keep between the Gulf Stream and the shoals, where they would have ``the advantage of the eddy current, which moves contrary to the Gulph Stream'' \cite{Franklin1786,RhinesHolland1979}. Eddies and the Stream have been linked ever since, but in what sense? In rotating turbulence, eddies can feed the jets that spawn them. Starr called this \emph{negative viscosity} \cite{Starr1968}: the eddy momentum flux runs up the gradient of the mean flow, into the jet. His evidence came from current-meter surveys of the Gulf Stream \cite{Webster1965}; once the stresses were rotated into the frame of the mean flow, R\"udiger found only positive viscosity \cite{Rudiger1976}. Both treated the eddy viscosity as a property of the flow; in Kraichnan's formulation of the same year it depends on the scale $k_c$ that separates mean flow from eddies, and can take either sign \cite{Kraichnan1976}. Starr went half way. Written as a closure, a negative viscosity is unstable: the energy it supplies grows with the energy of the flow, a positive feedback \cite{Sukorianskyetal1996}, and the idea has remained descriptive. Up-gradient eddy stresses have since been called ``antifrictional'' \cite{McIntyre2008,DritschelMcIntyre2008} or ``antidiffusive'' \cite{Hughesetal2026}, and the term itself has fallen out of use, as ``what used to be called negative viscosity'' \cite{McIntyre2008}. There is, however, nothing anti-diffusive in them: in two-dimensional turbulence the up-gradient flux is the large-scale signature of the inverse energy cascade. Nor are antifriction and antidiffusion interchangeable: on the $\beta$-plane, down-gradient mixing of potential vorticity drives an up-gradient flux of momentum \cite{RhinesHolland1979,DritschelMcIntyre2008}, so a negative viscosity coexists with a positive diffusivity.

Two-dimensional turbulence supplies the missing half. When energy is injected at small scales, the transfer to a resolved 
scale $k$ below a cutoff $k_c$ is a \emph{stabilized} negative viscosity (SNV) \cite{Sukorianskyetal1996,Chekhlovetal1996},
\begin{align}
&\nu(k|k_c)=-a\left[1-c\,(k/k_c)^2\right], \label{eq:snv} \\
&c=8/5\ \text{(asymptotic theory)}, \nonumber
\end{align}
in physical space a negative Laplacian balanced by a positive biharmonic term. It is obtained by truncating Kraichnan's two-parametric viscosity \cite{Kraichnan1976} to two terms, with $c$ fixed by the condition of zero enstrophy transfer in the energy subrange (for a Kolmogorov spectrum) and $a=25\Pi_\varepsilon/(18\bar\Omega)$ by the energy flux $\Pi_\varepsilon$ into the resolved scales, where $\bar\Omega$ is the resolved enstrophy; this dependence on the flow keeps the energy input constant and removes the feedback \cite{Sukorianskyetal1996}; the underlying $\nu(k|k_c)$ was confirmed in direct numerical simulation, in agreement with closure and renormalization-group theory \cite{Chekhlovetal1994}. The large scales are fed, the scales near the cutoff are damped, and the closure is well posed. On a $\beta$-plane the transfer becomes anisotropic and organizes the flow into zonal jets \cite{Galperinetal2004,Sukorianskyetal2007}.

Classical theory of the wind-driven ocean closes the western boundary current with friction: bottom drag in Stommel's model \cite{Stommel1948}, a positive lateral viscosity in Munk's \cite{Munk1950}. Western boundary currents are therefore a natural place to test whether geophysical jets obey SNV. The test requires eddy momentum fluxes resolved across a jet along its whole path, which the SWOT satellite now provides. Here we apply it to all five subtropical western boundary currents.

\textit{Measurement.}---We use SWOT KaRIn Level-2 sea-surface height (version D), Aug 2023--Jul 2025 (34--35 cycles per current), and compute geostrophic velocity anomalies in the swath, binned at $0.25^\circ$. The mean current---axis, direction and along-stream velocity $U_s(n)$---is taken from the GLORYS12 reanalysis \cite{Lelloucheetal2021} for the same period. Eddy stresses are rotated into the current's own frame $(s,n)$, as R\"udiger required \cite{Rudiger1976}, and averaged over 300-km blocks along the path. On each flank, 25--100~km from the axis,
\begin{equation}
\ve=-\langle u_s'u_n'\rangle\big/(\partial U_s/\partial n),
\label{eq:nue}
\end{equation}
positive for a down-gradient and negative for an up-gradient flux. Across the current we fit $\langle u_s'u_n'\rangle=A_2\,\partial_n U_s+A_4\,\partial_n^3 U_s$; the SNV structure corresponds to $A_2>0$, $A_4>0$. Uncertainties are 95\% intervals from bootstrap resampling of the satellite cycles. Details are given in the Supplemental Material \cite{SMsnv}.

\textit{Coastal currents and topographic $\beta$.}---Western boundary currents are coastal currents: upstream of separation they flow along the continental slope, where the mean PV gradient is set by topography. Along the slope the topographic $\beta$, $f|\nabla H|/H$, exceeds the planetary value by a factor of 30--115, and the currents run within 2--11$^\circ$ of the $f/H$ contours. At separation the topographic $\beta$ collapses by a factor of 10--50 within a few hundred kilometres, and the current crosses the $f/H$ contours at 19--37$^\circ$. In the free jets $\beta$ is again close to planetary ($\beta_t/\beta_p\approx1$--30). The three regimes of eddy transfer below coincide with these three regimes of $\beta$.

\begin{figure*}
\includegraphics[width=\textwidth]{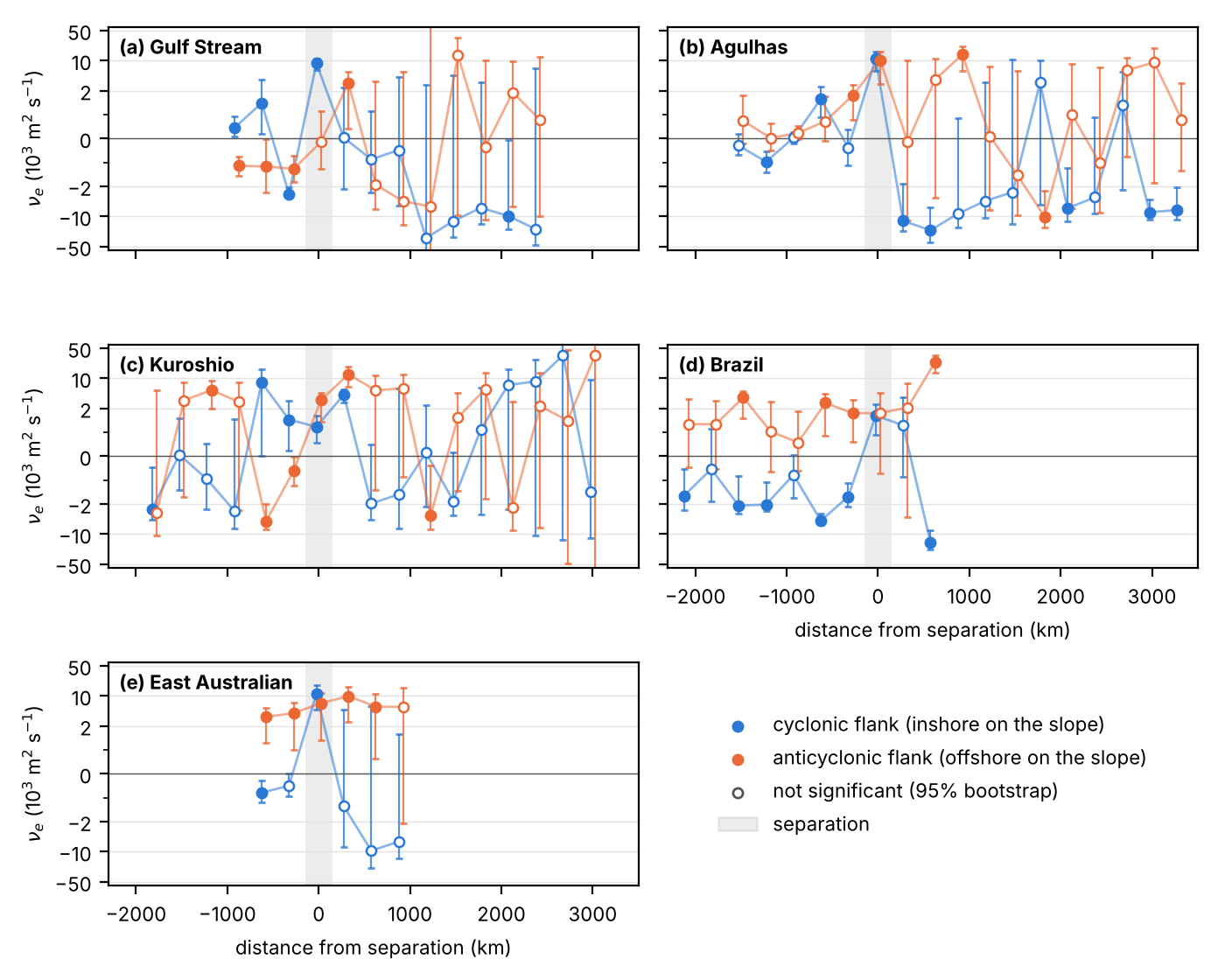}
\caption{Effective eddy viscosity (\ref{eq:nue}) from SWOT eddy fluxes (Aug 2023--Jul 2025) along the five subtropical western boundary currents, aligned at separation from the continental slope (gray band; negative distance = upstream, on the slope). Blue: cyclonic flank (inshore on the slope); orange: anticyclonic flank. Bars: 95\% bootstrap intervals over satellite cycles; open symbols: not significant. Symmetric-log axis, linear within $\pm2\times10^3$~m$^2$\,s$^{-1}$.}
\label{fig:five}
\end{figure*}

\textit{Three regimes.}---Figure~\ref{fig:five} shows $\ve$ along the five currents, aligned at the point where each leaves the continental slope. We label the flanks by the sense of the mean shear: the cyclonic flank is inshore on the slope.

\emph{On the slope} the eddies act with a negative viscosity of order $-10^3$~m$^2$\,s$^{-1}$: in the Gulf Stream $-(1.1\text{--}1.3)\times10^3$ on the offshore flank in every block, and $-3.0\,[-3.7,-2.1]\times10^3$ inshore at 33.8$^\circ$N; in the Agulhas $-1.0\,[-1.4,-0.5]\times10^3$; in the Brazil Current $-(1.7\text{--}4.9)\times10^3$ inshore in five of seven blocks. The SNV structure is resolved in the Gulf Stream ($A_2>0$, $A_4>0$ in all resamplings at 30$^\circ$N and 34$^\circ$N) and in the Agulhas ($A_4>0$ in all, $A_2>0$ in 93\%).

\emph{At separation} the sign reverses in all five currents. The cyclonic flank becomes positive: $+8.6\,[+6.1,+10.9]\times10^3$~m$^2$\,s$^{-1}$ at Cape Hatteras, $+11.0\,[+5.7,+15.9]\times10^3$ at the tip of the Agulhas Bank, $+11.1\,[+4.8,+18.0]\times10^3$ in the East Australian Current, and $+4.1\,[+2.7,+5.6]\times10^3$ (at 37.3$^\circ$N; $+1.2\times10^3$ at 35.1$^\circ$N) and $+1.7\,[+0.9,+2.5]\times10^3$ in the Kuroshio and Brazil Current. Here the fit gives $A_2<0$ in all resamplings: the eddies brake the current.

\emph{In the free jet} the negative viscosity returns on the cyclonic flank, ten times stronger than on the slope: $-13.8\,[-24.3,-3.5]\times10^3$~m$^2$\,s$^{-1}$ averaged over 1200--2400~km past Hatteras, and $-(6\text{--}21)\times10^3$ along the Agulhas retroflection and Return Current, where the SNV signs hold in all resamplings at 36--38$^\circ$E.

Two currents depart from this sequence, each for an identifiable reason. The Kuroshio was in its large-meander state during the whole record, and its upstream blocks are mixed. The East Australian Current is braked along its whole length; it barely forms a coherent jet before breaking into eddies.

\textit{A potential-vorticity test.}---In eddy-driven mean-flow theory the eddies mix potential vorticity down its gradient \cite{RhinesHolland1979}. On the slope, where that gradient is set by topographic $\beta$, a potential-vorticity diffusivity $\kappa$ gives, through the Taylor identity, $\langle u_s'u_n'\rangle\approx\kappa\beta_t n-\kappa\,\partial_nU_s$: a term linear in $n$, which could masquerade as negative viscosity in the fit, plus a term $-\kappa\,\partial_nU_s$ that the fit reads as a positive viscosity. Adding $A_0+A_1n$ to the fit (SWOT, 200 resamplings), the SNV signs ($A_2>0$, $A_4>0$) persist in nearly all resamplings at 29.9$^\circ$N in the Gulf Stream, where the Loop Current reattaches to the West Florida slope and in the Florida Straits, and in the Agulhas Return Current, and in most resamplings on the Agulhas slope (32.1$^\circ$S) and in the Kuroshio at 34.5$^\circ$N. At 33.8$^\circ$N in the Gulf Stream and where the Kuroshio leaves the East China Sea, the linear term and a positive viscosity account for the flux; in the Brazil and East Australian Currents the SNV structure is not resolved with or without these terms. The braking at separation survives in the Gulf Stream, Kuroshio and Agulhas. The up-gradient flux is therefore stabilized negative viscosity in some places and indistinguishable from down-gradient potential-vorticity mixing in others; the scale split, in which the small eddies feed the current almost everywhere, favors SNV, but the mean-flow fit alone does not settle it (Supplemental Material, S10).

\textit{Why the sign reverses.}---The sign follows the three regimes of $\beta$. In $\beta$-plane turbulence it is $\beta$ that orders the inverse transfer: Rossby-wave detuning of the triad interactions suppresses the negative viscosity for wavevectors away from the zonal direction, more sharply the stronger $\beta$ is: in second-order spectral closures the strength of a triad interaction is proportional to its relaxation time \cite{Orszag1977}, which depends on $\beta$ and reorients the inverse cascade toward zonal modes \cite{Huangetal2001}. A current aligned with strong topographic $\beta$ is therefore fed; where it leaves that constraint, the feeding is switched off and the eddies drain it. In the free jet, $\beta$ is planetary and the current is again nearly zonal, and the feeding returns. That the fed jets are eastward is consistent with the asymmetry of zonostrophic turbulence, in which eastward jets are sharp and stable and westward jets shed eddies \cite{Cohenetal2024}. R\"udiger noted that the anisotropy imposed by the coast lay outside his analysis \cite{Rudiger1976}; in our data it is this anisotropy, through topographic $\beta$, that decides the sign, and his positive viscosity is what we find where its constraint is lost.

\begin{figure}
\includegraphics[width=\columnwidth]{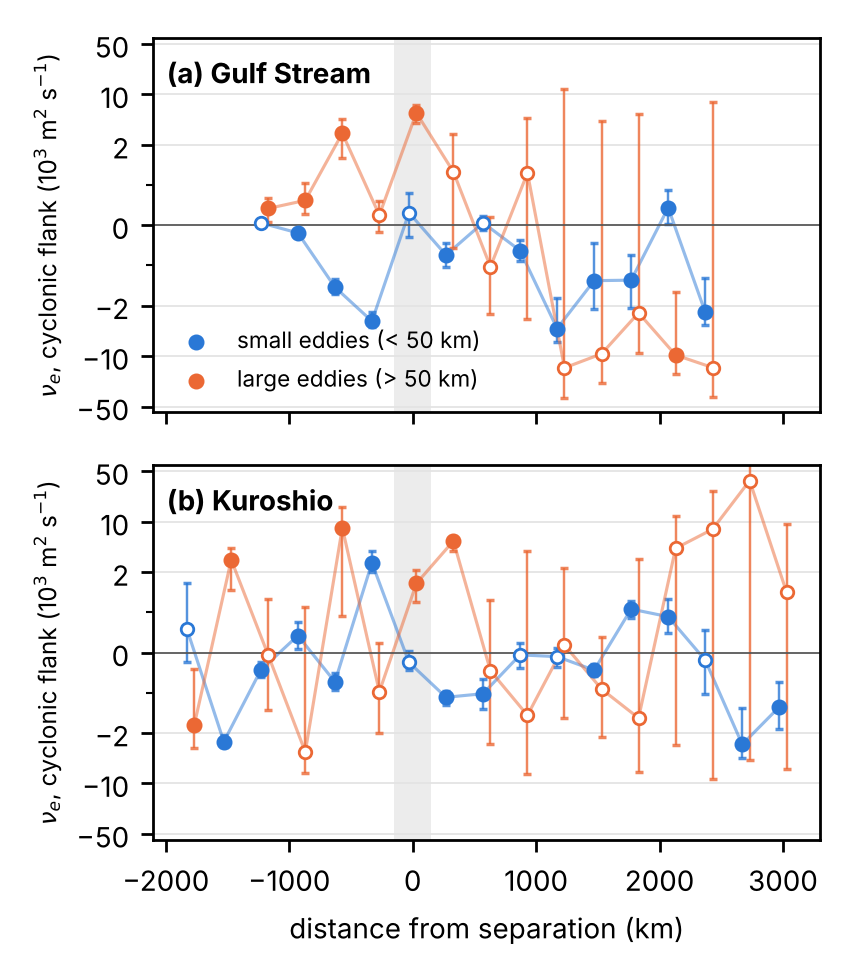}
\caption{Scale split of the SWOT eddy viscosity on the cyclonic flank (along-track Gaussian filter, $\ell=50$~km): small eddies (blue) and large eddies (orange), aligned at separation (gray band). Bars: 95\% bootstrap intervals; open symbols: not significant.}
\label{fig:split}
\end{figure}

\textit{Which eddies brake.}---Splitting the SWOT eddy flux at 50~km along track resolves the braking. Small eddies keep feeding the jet on the cyclonic flank, significantly in 18 of 30 blocks in the Gulf Stream and Kuroshio. The braking comes from the large eddies, with scales comparable to the jet: at Cape Hatteras they contribute $+5.6\times10^3$~m$^2$\,s$^{-1}$ against $+0.3\times10^3$ from small eddies, and at the Kuroshio separation $+5.4\times10^3$ against $-1.1\times10^3$. In SNV terms, the collapse of topographic $\beta$ lets the eddies grow to the jet scale, where the transfer enters the positive cusp of $\nu(k|k_c)$ near the cutoff: the same closure that feeds the jet upstream brakes it at separation (Fig.~\ref{fig:split}). In the ocean, moreover, there is no steady state: the flux $\Pi_\varepsilon$ across the cutoff is not imposed but delivered by the flow, so the SNV derivation holds instant by instant, with an amplitude $a=25\Pi_\varepsilon/(18\bar\Omega)$ that follows the flux. In GLORYS the measured $a$ tracks $\Pi_\varepsilon$ month by month in the Gulf Stream, but not $1/\bar\Omega$ (Supplemental Material, S7). Where the inverse transfer is interrupted and $\Pi_\varepsilon$ reverses, the same closure yields a positive Laplacian viscosity: the sign reversal at separation is then the closure's own prediction. Measuring the sign of $\Pi_\varepsilon$ at jet scales along the path is a direct test.

\textit{Related formulations.}---Anti-diffusive transport was recently surveyed across disciplines \cite{Hughesetal2026}; momentum anti-diffusion has been derived perturbatively for two-dimensional condensates \cite{Frishmanetal2026} and invoked for superrotation \cite{Vallisetal2026}, but without a scale- and regime-dependent formulation. For parallel mean flows the Taylor identity \cite{Taylor1915}, $-\partial_y\langle u'v'\rangle=\langle v'\zeta'\rangle$, is exact, but it does not fix the closure. Down-gradient potential-vorticity mixing, as in potential-vorticity-staircase mechanisms \cite{DritschelMcIntyre2008}, removes energy and enstrophy from the resolved flow together, whereas in two-dimensional turbulence the dual cascade supplies energy to the large scales while sending enstrophy to small ones. SNV satisfies both constraints, its shape being fixed by zero net enstrophy transfer \cite{Sukorianskyetal1996}. On the $\beta$-plane the zonal $k^{-5}$ spectrum implies enstrophy accumulating in the jets, a further feature that down-gradient mixing does not describe. The same operator, a negative Laplacian stabilized by a positive biharmonic, has arisen independently with constant coefficients, in spinodal decomposition (Cahn--Hilliard \cite{CahnHilliard1958}) and in flame fronts and chemical oscillations (Kuramoto--Sivashinsky \cite{KuramotoTsuzuki1976,Sivashinsky1977}); there the operator is fixed, and a constant-coefficient version for two-dimensional flow failed to reproduce even the Kolmogorov spectrum \cite{Gamaetal1991}. Subgrid backscatter has also been represented stochastically \cite{Leith1990} and, in eddy-permitting ocean models, deterministically, as a negative Laplacian viscosity with biharmonic dissipation whose amplitude is set by a subgrid energy budget \cite{JansenHeld2014,Jansenetal2015}. In SNV both coefficients are scale- and flow-dependent properties of turbulent transfer relative to the cutoff $k_c$: negative eddy viscosity is a property of two-dimensional turbulence, not of the operator's form. It is also a property of the resolution, not of the fluid: in a simulation resolving all scales the eddy viscosity vanishes and only the positive molecular viscosity remains. The sign reversals reported here are therefore properties of the flow coarse-grained at the scales SWOT resolves. Any positive higher-order term would regularize the closure; the biharmonic is the form selected by the derived $\nu(k|k_c)$ and confirmed in our simulations.

\begin{figure}
\includegraphics[width=\columnwidth]{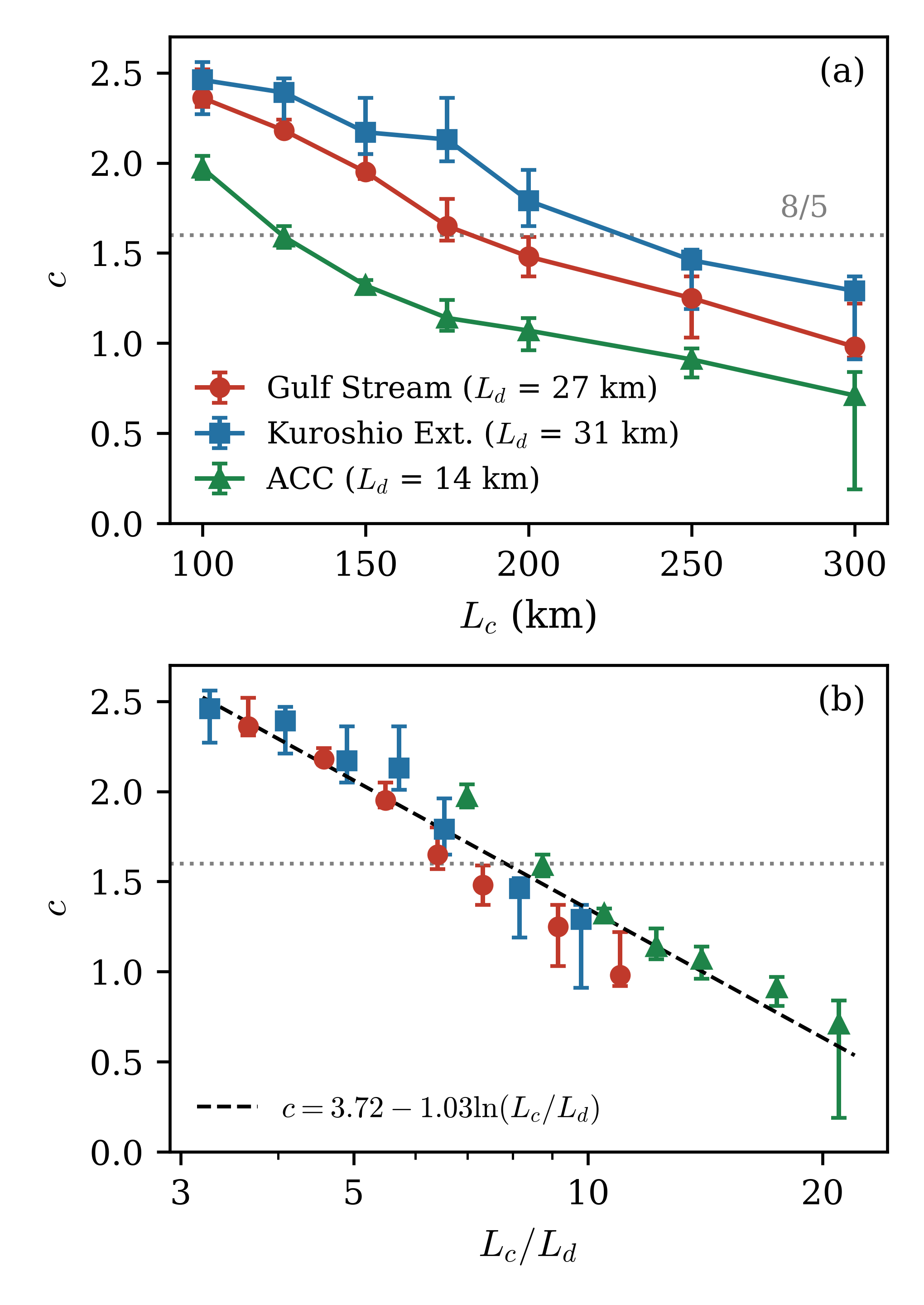}
\caption{SNV parameter $c$ measured in GLORYS12 daily surface currents (2015) in three basins, (a) versus the cutoff $L_c$ and (b) versus $L_c/L_d$; bars: interquartile range over months. Dotted: $c=8/5$ of the asymptotic theory; dashed: fit $c=3.72-1.03\ln(L_c/L_d)$, $R^2=0.94$.}
\label{fig:snv}
\end{figure}

\textit{The SNV form.}---The structure of Eq.~(\ref{eq:snv}) can be tested spectrally in eddy-resolving data. In GLORYS daily surface currents for the Gulf Stream, Kuroshio Extension and Antarctic Circumpolar Current, the two-parametric viscosity, averaged over the directions of the wavevector, has $a>0$ in all 48 months, and $c$ depends on the cutoff only through $L_c/L_d$: $c=3.72-1.03\ln(L_c/L_d)$ ($R^2=0.94$), passing through $8/5$ at $L_c\approx8L_d$ (Fig.~\ref{fig:snv}). This is expected: the value $8/5$ follows from a Kolmogorov spectrum near the cutoff, with forcing deep in the subgrid range, whereas the ocean injects energy near $L_d$, so $c$ approaches $8/5$ only when $L_c\gg L_d$. The deformation radius $L_d$, the scale of baroclinic energy injection, thus plays the role of the forcing scale of two-dimensional SNV. This is consistent with the reversal of the spectral energy flux at a scale set by $L_d$ in laboratory baroclinic turbulence \cite{Dingetal2026} and on Jupiter \cite{YoungRead2017}. On the $\beta$-plane the viscosity is anisotropic, negative near the zonal wavevectors and vanishing elsewhere \cite{Chekhlovetal1996}; resolving this anisotropy in the ocean requires higher-resolution data over larger domains and is a direct test for the future.

\textit{Discussion.}---In the western boundary currents the eddies do not act as Munk's friction. On the slope, where Munk's layer requires a positive lateral viscosity of $10^3$--$10^4$~m$^2$\,s$^{-1}$, we measure a negative one of order $10^3$. Downstream the eddies feed the free jets with a negative viscosity of order $10^4$. The only place where they act as friction is separation, where the constraint that orders the eddy transfer is lost. The vorticity that the wind supplies to the gyre must still be removed. Our measurements suggest that the sink which classical theory spreads along the whole boundary layer is provided on the slope by bottom drag and form stress, which SWOT does not see, and is concentrated by the eddies at separation. In each negative regime the transfer is up-gradient in momentum, and where resolved it has the stabilized form, so the flux does not run away to small scales. Negative viscosity is neither antidiffusion nor a misleading notion; with its stabilization it is the closure that turbulence theory predicts and the ocean obeys.

Two limitations: SWOT gives the horizontal surface flux only, not vertical transfer by form stress or by submesoscale instabilities, which on the North Wall in winter extract kinetic energy from the current \cite{Thomasetal2013}; and the mean shear comes from a reanalysis. In eddy-resolving models the separation and penetration of the Gulf Stream depend on the choice between Laplacian and biharmonic viscosity, and the best results combine the two with positive coefficients \cite{ChassignetGarraffo2001}; in the ocean, the Laplacian coefficient supplied by the eddies changes sign with the regime of $\beta$. Applying SNV to ocean models has been held back by two unknowns: the energy flux $\Pi_\varepsilon$ into the resolved scales and the enstrophy $\bar\Omega$ that normalizes it, which the closure does not fix. Eddy-resolving data now provide $\Pi_\varepsilon$ directly, and it, rather than $\bar\Omega$, controls the amplitude.

Einstein, whose assistant Kraichnan went on to found the field-theoretical approach to turbulence from which SNV descends \cite{GalperinSukoriansky2024}, put it best: ``Raffiniert ist der Herrgott, aber boshaft ist er nicht'' (``Subtle is the Lord, but malicious He is not'') \cite{Pais1982}. The instability that would ruin a negative-viscosity closure is tamed twice: by the flow, which sets the strength of the negative viscosity, and by a positive biharmonic eddy viscosity, which the ocean shows.

\textit{Data availability.}---SWOT and GLORYS12 data are publicly available; the scripts and derived block statistics are described in the Supplemental Material \cite{SMsnv}.

\begin{acknowledgments}
We dedicate this Letter to the memory of Steven A.~Orszag and Victor Yakhot, with whom this line of work began in Princeton. SWOT data: NASA/CNES via PO.DAAC; GLORYS12: Copernicus Marine Service. Data analysis and drafting were assisted by Claude (Anthropic), an AI system; the authors take full responsibility for the content.
\end{acknowledgments}


\bibliographystyle{apsrev4-2}
\bibliography{HurricaneBib}

\begin{thebibliography}{37}%
\makeatletter
\providecommand \@ifxundefined [1]{%
 \@ifx{#1\undefined}
}%
\providecommand \@ifnum [1]{%
 \ifnum #1\expandafter \@firstoftwo
 \else \expandafter \@secondoftwo
 \fi
}%
\providecommand \@ifx [1]{%
 \ifx #1\expandafter \@firstoftwo
 \else \expandafter \@secondoftwo
 \fi
}%
\providecommand \natexlab [1]{#1}%
\providecommand \enquote  [1]{``#1''}%
\providecommand \bibnamefont  [1]{#1}%
\providecommand \bibfnamefont [1]{#1}%
\providecommand \citenamefont [1]{#1}%
\providecommand \href@noop [0]{\@secondoftwo}%
\providecommand \href [0]{\begingroup \@sanitize@url \@href}%
\providecommand \@href[1]{\@@startlink{#1}\@@href}%
\providecommand \@@href[1]{\endgroup#1\@@endlink}%
\providecommand \@sanitize@url [0]{\catcode `\\12\catcode `\$12\catcode
  `\&12\catcode `\#12\catcode `\^12\catcode `\_12\catcode `\%12\relax}%
\providecommand \@@startlink[1]{}%
\providecommand \@@endlink[0]{}%
\providecommand \url  [0]{\begingroup\@sanitize@url \@url }%
\providecommand \@url [1]{\endgroup\@href {#1}{\urlprefix }}%
\providecommand \urlprefix  [0]{URL }%
\providecommand \Eprint [0]{\href }%
\providecommand \doibase [0]{https://doi.org/}%
\providecommand \selectlanguage [0]{\@gobble}%
\providecommand \bibinfo  [0]{\@secondoftwo}%
\providecommand \bibfield  [0]{\@secondoftwo}%
\providecommand \translation [1]{[#1]}%
\providecommand \BibitemOpen [0]{}%
\providecommand \bibitemStop [0]{}%
\providecommand \bibitemNoStop [0]{.\EOS\space}%
\providecommand \EOS [0]{\spacefactor3000\relax}%
\providecommand \BibitemShut  [1]{\csname bibitem#1\endcsname}%
\let\auto@bib@innerbib\@empty
\bibitem [{\citenamefont {Franklin}(1786)}]{Franklin1786}%
  \BibitemOpen
  \bibfield  {author} {\bibinfo {author} {\bibfnamefont {B.}~\bibnamefont
  {Franklin}},\ }\href@noop {} {\bibfield  {journal} {\bibinfo  {journal}
  {Trans. Am. Philos. Soc.}\ }\textbf {\bibinfo {volume} {2}},\ \bibinfo
  {pages} {294} (\bibinfo {year} {1786})}\BibitemShut {NoStop}%
\bibitem [{\citenamefont {Rhines}\ and\ \citenamefont
  {Holland}(1979)}]{RhinesHolland1979}%
  \BibitemOpen
  \bibfield  {author} {\bibinfo {author} {\bibfnamefont {P.~B.}\ \bibnamefont
  {Rhines}}\ and\ \bibinfo {author} {\bibfnamefont {W.~R.}\ \bibnamefont
  {Holland}},\ }\href@noop {} {\bibfield  {journal} {\bibinfo  {journal} {Dyn.
  Atmos. Oceans}\ }\textbf {\bibinfo {volume} {3}},\ \bibinfo {pages} {289}
  (\bibinfo {year} {1979})}\BibitemShut {NoStop}%
\bibitem [{\citenamefont {Starr}(1968)}]{Starr1968}%
  \BibitemOpen
  \bibfield  {author} {\bibinfo {author} {\bibfnamefont {V.~P.}\ \bibnamefont
  {Starr}},\ }\href@noop {} {\emph {\bibinfo {title} {Physics of Negative
  Viscosity Phenomena}}}\ (\bibinfo  {publisher} {McGraw-Hill},\ \bibinfo
  {address} {New York},\ \bibinfo {year} {1968})\BibitemShut {NoStop}%
\bibitem [{\citenamefont {Webster}(1965)}]{Webster1965}%
  \BibitemOpen
  \bibfield  {author} {\bibinfo {author} {\bibfnamefont {F.}~\bibnamefont
  {Webster}},\ }\href@noop {} {\bibfield  {journal} {\bibinfo  {journal}
  {Tellus}\ }\textbf {\bibinfo {volume} {17}},\ \bibinfo {pages} {239}
  (\bibinfo {year} {1965})}\BibitemShut {NoStop}%
\bibitem [{\citenamefont {R{\"u}diger}(1976)}]{Rudiger1976}%
  \BibitemOpen
  \bibfield  {author} {\bibinfo {author} {\bibfnamefont {G.}~\bibnamefont
  {R{\"u}diger}},\ }\href@noop {} {\bibfield  {journal} {\bibinfo  {journal}
  {Tellus}\ }\textbf {\bibinfo {volume} {28}},\ \bibinfo {pages} {183}
  (\bibinfo {year} {1976})}\BibitemShut {NoStop}%
\bibitem [{\citenamefont {Kraichnan}(1976)}]{Kraichnan1976}%
  \BibitemOpen
  \bibfield  {author} {\bibinfo {author} {\bibfnamefont {R.~H.}\ \bibnamefont
  {Kraichnan}},\ }\href@noop {} {\bibfield  {journal} {\bibinfo  {journal} {J.
  Atmos. Sci.}\ }\textbf {\bibinfo {volume} {33}},\ \bibinfo {pages} {1521}
  (\bibinfo {year} {1976})}\BibitemShut {NoStop}%
\bibitem [{\citenamefont {Sukoriansky}\ \emph {et~al.}(1996)\citenamefont
  {Sukoriansky}, \citenamefont {Chekhlov}, \citenamefont {Orszag},
  \citenamefont {Galperin},\ and\ \citenamefont
  {Staroselsky}}]{Sukorianskyetal1996}%
  \BibitemOpen
  \bibfield  {author} {\bibinfo {author} {\bibfnamefont {S.}~\bibnamefont
  {Sukoriansky}}, \bibinfo {author} {\bibfnamefont {A.}~\bibnamefont
  {Chekhlov}}, \bibinfo {author} {\bibfnamefont {S.~A.}\ \bibnamefont
  {Orszag}}, \bibinfo {author} {\bibfnamefont {B.}~\bibnamefont {Galperin}},\
  and\ \bibinfo {author} {\bibfnamefont {I.}~\bibnamefont {Staroselsky}},\
  }\href@noop {} {\bibfield  {journal} {\bibinfo  {journal} {J. Sci. Comput.}\
  }\textbf {\bibinfo {volume} {11}},\ \bibinfo {pages} {13} (\bibinfo {year}
  {1996})}\BibitemShut {NoStop}%
\bibitem [{\citenamefont {McIntyre}(2008)}]{McIntyre2008}%
  \BibitemOpen
  \bibfield  {author} {\bibinfo {author} {\bibfnamefont {M.~E.}\ \bibnamefont
  {McIntyre}},\ }\href@noop {} {\bibfield  {journal} {\bibinfo  {journal} {Adv.
  Geosci.}\ }\textbf {\bibinfo {volume} {15}},\ \bibinfo {pages} {47} (\bibinfo
  {year} {2008})}\BibitemShut {NoStop}%
\bibitem [{\citenamefont {Dritschel}\ and\ \citenamefont
  {McIntyre}(2008)}]{DritschelMcIntyre2008}%
  \BibitemOpen
  \bibfield  {author} {\bibinfo {author} {\bibfnamefont {D.~G.}\ \bibnamefont
  {Dritschel}}\ and\ \bibinfo {author} {\bibfnamefont {M.~E.}\ \bibnamefont
  {McIntyre}},\ }\href@noop {} {\bibfield  {journal} {\bibinfo  {journal} {J.
  Atmos. Sci.}\ }\textbf {\bibinfo {volume} {65}},\ \bibinfo {pages} {855}
  (\bibinfo {year} {2008})}\BibitemShut {NoStop}%
\bibitem [{\citenamefont {Hughes}\ \emph {et~al.}(2026)\citenamefont {Hughes},
  \citenamefont {Diamond}, \citenamefont {Frishman},\ and\ \citenamefont
  {Pandit}}]{Hughesetal2026}%
  \BibitemOpen
  \bibfield  {author} {\bibinfo {author} {\bibfnamefont {D.~W.}\ \bibnamefont
  {Hughes}}, \bibinfo {author} {\bibfnamefont {P.~H.}\ \bibnamefont {Diamond}},
  \bibinfo {author} {\bibfnamefont {A.}~\bibnamefont {Frishman}},\ and\
  \bibinfo {author} {\bibfnamefont {R.}~\bibnamefont {Pandit}},\ }\href@noop {}
  {\bibfield  {journal} {\bibinfo  {journal} {Phil. Trans. R. Soc. A}\ }\textbf
  {\bibinfo {volume} {384}},\ \bibinfo {eid} {20250132} (\bibinfo {year}
  {2026})}\BibitemShut {NoStop}%
\bibitem [{\citenamefont {Chekhlov}\ \emph {et~al.}(1996)\citenamefont
  {Chekhlov}, \citenamefont {Orszag}, \citenamefont {Sukoriansky},
  \citenamefont {Galperin},\ and\ \citenamefont
  {Staroselsky}}]{Chekhlovetal1996}%
  \BibitemOpen
  \bibfield  {author} {\bibinfo {author} {\bibfnamefont {A.}~\bibnamefont
  {Chekhlov}}, \bibinfo {author} {\bibfnamefont {S.~A.}\ \bibnamefont
  {Orszag}}, \bibinfo {author} {\bibfnamefont {S.}~\bibnamefont {Sukoriansky}},
  \bibinfo {author} {\bibfnamefont {B.}~\bibnamefont {Galperin}},\ and\
  \bibinfo {author} {\bibfnamefont {I.}~\bibnamefont {Staroselsky}},\ }\href
  {https://doi.org/10.1016/0167-2789(96)00102-9} {\bibfield
  {journal} {\bibinfo  {journal} {Phys. D}\ }\textbf {\bibinfo {volume} {98}},\
  \bibinfo {pages} {321} (\bibinfo {year} {1996})}\BibitemShut {NoStop}%
\bibitem [{\citenamefont {Chekhlov}\ \emph {et~al.}(1994)\citenamefont
  {Chekhlov}, \citenamefont {Orszag}, \citenamefont {Sukoriansky},
  \citenamefont {Galperin},\ and\ \citenamefont
  {Staroselsky}}]{Chekhlovetal1994}%
  \BibitemOpen
  \bibfield  {author} {\bibinfo {author} {\bibfnamefont {A.}~\bibnamefont
  {Chekhlov}}, \bibinfo {author} {\bibfnamefont {S.~A.}\ \bibnamefont
  {Orszag}}, \bibinfo {author} {\bibfnamefont {S.}~\bibnamefont {Sukoriansky}},
  \bibinfo {author} {\bibfnamefont {B.}~\bibnamefont {Galperin}},\ and\
  \bibinfo {author} {\bibfnamefont {I.}~\bibnamefont {Staroselsky}},\
  }\href@noop {} {\bibfield  {journal} {\bibinfo  {journal} {Phys. Fluids}\
  }\textbf {\bibinfo {volume} {6}},\ \bibinfo {pages} {2548} (\bibinfo {year}
  {1994})}\BibitemShut {NoStop}%
\bibitem [{\citenamefont {Galperin}\ \emph {et~al.}(2004)\citenamefont
  {Galperin}, \citenamefont {Nakano}, \citenamefont {Huang},\ and\
  \citenamefont {Sukoriansky}}]{Galperinetal2004}%
  \BibitemOpen
  \bibfield  {author} {\bibinfo {author} {\bibfnamefont {B.}~\bibnamefont
  {Galperin}}, \bibinfo {author} {\bibfnamefont {H.}~\bibnamefont {Nakano}},
  \bibinfo {author} {\bibfnamefont {H.-P.}\ \bibnamefont {Huang}},\ and\
  \bibinfo {author} {\bibfnamefont {S.}~\bibnamefont {Sukoriansky}},\
  }\href@noop {} {\bibfield  {journal} {\bibinfo  {journal} {Geophys. Res.
  Lett.}\ }\textbf {\bibinfo {volume} {31}},\ \bibinfo {eid} {L13303} (\bibinfo
  {year} {2004})}\BibitemShut {NoStop}%
\bibitem [{\citenamefont {Sukoriansky}\ \emph {et~al.}(2007)\citenamefont
  {Sukoriansky}, \citenamefont {Dikovskaya},\ and\ \citenamefont
  {Galperin}}]{Sukorianskyetal2007}%
  \BibitemOpen
  \bibfield  {author} {\bibinfo {author} {\bibfnamefont {S.}~\bibnamefont
  {Sukoriansky}}, \bibinfo {author} {\bibfnamefont {N.}~\bibnamefont
  {Dikovskaya}},\ and\ \bibinfo {author} {\bibfnamefont {B.}~\bibnamefont
  {Galperin}},\ }\href@noop {} {\bibfield  {journal} {\bibinfo  {journal} {J.
  Atmos. Sci.}\ }\textbf {\bibinfo {volume} {64}},\ \bibinfo {pages} {3312}
  (\bibinfo {year} {2007})}\BibitemShut {NoStop}%
\bibitem [{\citenamefont {Stommel}(1948)}]{Stommel1948}%
  \BibitemOpen
  \bibfield  {author} {\bibinfo {author} {\bibfnamefont {H.}~\bibnamefont
  {Stommel}},\ }\href@noop {} {\bibfield  {journal} {\bibinfo  {journal}
  {Trans. Am. Geophys. Union}\ }\textbf {\bibinfo {volume} {29}},\ \bibinfo
  {pages} {202} (\bibinfo {year} {1948})}\BibitemShut {NoStop}%
\bibitem [{\citenamefont {Munk}(1950)}]{Munk1950}%
  \BibitemOpen
  \bibfield  {author} {\bibinfo {author} {\bibfnamefont {W.~H.}\ \bibnamefont
  {Munk}},\ }\href@noop {} {\bibfield  {journal} {\bibinfo  {journal} {J.
  Meteorol.}\ }\textbf {\bibinfo {volume} {7}},\ \bibinfo {pages} {79}
  (\bibinfo {year} {1950})}\BibitemShut {NoStop}%
\bibitem [{\citenamefont {Lellouche}\ \emph {et~al.}(2021)\citenamefont
  {Lellouche} \emph {et~al.}}]{Lelloucheetal2021}%
  \BibitemOpen
  \bibfield  {author} {\bibinfo {author} {\bibfnamefont {J.-M.}\ \bibnamefont
  {Lellouche}} \emph {et~al.},\ }\href@noop {} {\bibfield  {journal} {\bibinfo
  {journal} {Front. Earth Sci.}\ }\textbf {\bibinfo {volume} {9}},\ \bibinfo
  {eid} {698876} (\bibinfo {year} {2021})}\BibitemShut {NoStop}%
\bibitem [{SMs()}]{SMsnv}%
  \BibitemOpen
  \href@noop {} {}\bibinfo {note} {See Supplemental Material at [URL will be
  inserted by publisher] for data processing, the current-following frame,
  block-by-block results, topographic $\beta$, the scale split, the spectral
  test of the SNV form and the potential-vorticity mixing test.}\BibitemShut
  {Stop}%
\bibitem [{\citenamefont {Orszag}(1977)}]{Orszag1977}%
  \BibitemOpen
  \bibfield  {author} {\bibinfo {author} {\bibfnamefont {S.~A.}\ \bibnamefont
  {Orszag}},\ }in\ \href@noop {} {\emph {\bibinfo {booktitle} {Fluid Dynamics,
  Les Houches Summer School in Physics 1973}}},\ \bibinfo {editor} {edited by\
  \bibinfo {editor} {\bibfnamefont {R.}~\bibnamefont {Balian}}\ and\ \bibinfo
  {editor} {\bibfnamefont {J.-L.}\ \bibnamefont {Peube}}}\ (\bibinfo
  {publisher} {Gordon and Breach},\ \bibinfo {address} {New York},\ \bibinfo
  {year} {1977})\ pp.\ \bibinfo {pages} {237--374}\BibitemShut {NoStop}%
\bibitem [{\citenamefont {Huang}\ \emph {et~al.}(2001)\citenamefont {Huang},
  \citenamefont {Galperin},\ and\ \citenamefont {Sukoriansky}}]{Huangetal2001}%
  \BibitemOpen
  \bibfield  {author} {\bibinfo {author} {\bibfnamefont {H.-P.}\ \bibnamefont
  {Huang}}, \bibinfo {author} {\bibfnamefont {B.}~\bibnamefont {Galperin}},\
  and\ \bibinfo {author} {\bibfnamefont {S.}~\bibnamefont {Sukoriansky}},\
  }\href@noop {} {\bibfield  {journal} {\bibinfo  {journal} {Phys. Fluids}\
  }\textbf {\bibinfo {volume} {13}},\ \bibinfo {pages} {225} (\bibinfo {year}
  {2001})}\BibitemShut {NoStop}%
\bibitem [{\citenamefont {Cohen}\ \emph {et~al.}(2024)\citenamefont {Cohen},
  \citenamefont {Galperin},\ and\ \citenamefont {Sukoriansky}}]{Cohenetal2024}%
  \BibitemOpen
  \bibfield  {author} {\bibinfo {author} {\bibfnamefont {N.}~\bibnamefont
  {Cohen}}, \bibinfo {author} {\bibfnamefont {B.}~\bibnamefont {Galperin}},\
  and\ \bibinfo {author} {\bibfnamefont {S.}~\bibnamefont {Sukoriansky}},\
  }\href {https://doi.org/10.3390/atmos15060711} {\bibfield  {journal}
  {\bibinfo  {journal} {MDPI Atmosphere}\ }\textbf {\bibinfo {volume} {15}},\
  \bibinfo {eid} {711} (\bibinfo {year} {2024})}\BibitemShut {NoStop}%
\bibitem [{\citenamefont {Frishman}\ \emph {et~al.}(2026)\citenamefont
  {Frishman}, \citenamefont {Gom{\'e}},\ and\ \citenamefont
  {Svirsky}}]{Frishmanetal2026}%
  \BibitemOpen
  \bibfield  {author} {\bibinfo {author} {\bibfnamefont {A.}~\bibnamefont
  {Frishman}}, \bibinfo {author} {\bibfnamefont {S.}~\bibnamefont {Gom{\'e}}},\
  and\ \bibinfo {author} {\bibfnamefont {A.}~\bibnamefont {Svirsky}},\
  }\href@noop {} {\bibfield  {journal} {\bibinfo  {journal} {Phil. Trans. R.
  Soc. A}\ }\textbf {\bibinfo {volume} {384}},\ \bibinfo {eid} {20250133}
  (\bibinfo {year} {2026})}\BibitemShut {NoStop}%
\bibitem [{\citenamefont {Vallis}\ \emph {et~al.}(2026)\citenamefont {Vallis},
  \citenamefont {Matilsky},\ and\ \citenamefont {Nicholas}}]{Vallisetal2026}%
  \BibitemOpen
  \bibfield  {author} {\bibinfo {author} {\bibfnamefont {G.~K.}\ \bibnamefont
  {Vallis}}, \bibinfo {author} {\bibfnamefont {L.}~\bibnamefont {Matilsky}},\
  and\ \bibinfo {author} {\bibfnamefont {Q.}~\bibnamefont {Nicholas}},\
  }\href@noop {} {\bibfield  {journal} {\bibinfo  {journal} {Phil. Trans. R.
  Soc. A}\ }\textbf {\bibinfo {volume} {384}},\ \bibinfo {eid} {20250135}
  (\bibinfo {year} {2026})}\BibitemShut {NoStop}%
\bibitem [{\citenamefont {Taylor}(1915)}]{Taylor1915}%
  \BibitemOpen
  \bibfield  {author} {\bibinfo {author} {\bibfnamefont {G.~I.}\ \bibnamefont
  {Taylor}},\ }\href@noop {} {\bibfield  {journal} {\bibinfo  {journal} {Phil.
  Trans. R. Soc. Lond. A}\ }\textbf {\bibinfo {volume} {215}},\ \bibinfo
  {pages} {1} (\bibinfo {year} {1915})}\BibitemShut {NoStop}%
\bibitem [{\citenamefont {Cahn}\ and\ \citenamefont
  {Hilliard}(1958)}]{CahnHilliard1958}%
  \BibitemOpen
  \bibfield  {author} {\bibinfo {author} {\bibfnamefont {J.~W.}\ \bibnamefont
  {Cahn}}\ and\ \bibinfo {author} {\bibfnamefont {J.~E.}\ \bibnamefont
  {Hilliard}},\ }\href@noop {} {\bibfield  {journal} {\bibinfo  {journal} {J.
  Chem. Phys.}\ }\textbf {\bibinfo {volume} {28}},\ \bibinfo {pages} {258}
  (\bibinfo {year} {1958})}\BibitemShut {NoStop}%
\bibitem [{\citenamefont {Kuramoto}\ and\ \citenamefont
  {Tsuzuki}(1976)}]{KuramotoTsuzuki1976}%
  \BibitemOpen
  \bibfield  {author} {\bibinfo {author} {\bibfnamefont {Y.}~\bibnamefont
  {Kuramoto}}\ and\ \bibinfo {author} {\bibfnamefont {T.}~\bibnamefont
  {Tsuzuki}},\ }\href@noop {} {\bibfield  {journal} {\bibinfo  {journal} {Prog.
  Theor. Phys.}\ }\textbf {\bibinfo {volume} {55}},\ \bibinfo {pages} {356}
  (\bibinfo {year} {1976})}\BibitemShut {NoStop}%
\bibitem [{\citenamefont {Sivashinsky}(1977)}]{Sivashinsky1977}%
  \BibitemOpen
  \bibfield  {author} {\bibinfo {author} {\bibfnamefont {G.~I.}\ \bibnamefont
  {Sivashinsky}},\ }\href@noop {} {\bibfield  {journal} {\bibinfo  {journal}
  {Acta Astronaut.}\ }\textbf {\bibinfo {volume} {4}},\ \bibinfo {pages} {1177}
  (\bibinfo {year} {1977})}\BibitemShut {NoStop}%
\bibitem [{\citenamefont {Gama}\ \emph {et~al.}(1991)\citenamefont {Gama},
  \citenamefont {Frisch},\ and\ \citenamefont {Scholl}}]{Gamaetal1991}%
  \BibitemOpen
  \bibfield  {author} {\bibinfo {author} {\bibfnamefont {S.}~\bibnamefont
  {Gama}}, \bibinfo {author} {\bibfnamefont {U.}~\bibnamefont {Frisch}},\ and\
  \bibinfo {author} {\bibfnamefont {H.}~\bibnamefont {Scholl}},\ }\href@noop {}
  {\bibfield  {journal} {\bibinfo  {journal} {J. Sci. Comput.}\ }\textbf
  {\bibinfo {volume} {6}},\ \bibinfo {pages} {425} (\bibinfo {year}
  {1991})}\BibitemShut {NoStop}%
\bibitem [{\citenamefont {Leith}(1990)}]{Leith1990}%
  \BibitemOpen
  \bibfield  {author} {\bibinfo {author} {\bibfnamefont {C.~E.}\ \bibnamefont
  {Leith}},\ }\href@noop {} {\bibfield  {journal} {\bibinfo  {journal} {Phys.
  Fluids A}\ }\textbf {\bibinfo {volume} {2}},\ \bibinfo {pages} {297}
  (\bibinfo {year} {1990})}\BibitemShut {NoStop}%
\bibitem [{\citenamefont {Jansen}\ and\ \citenamefont
  {Held}(2014)}]{JansenHeld2014}%
  \BibitemOpen
  \bibfield  {author} {\bibinfo {author} {\bibfnamefont {M.~F.}\ \bibnamefont
  {Jansen}}\ and\ \bibinfo {author} {\bibfnamefont {I.~M.}\ \bibnamefont
  {Held}},\ }\href@noop {} {\bibfield  {journal} {\bibinfo  {journal} {Ocean
  Modell.}\ }\textbf {\bibinfo {volume} {80}},\ \bibinfo {pages} {36} (\bibinfo
  {year} {2014})}\BibitemShut {NoStop}%
\bibitem [{\citenamefont {Jansen}\ \emph {et~al.}(2015)\citenamefont {Jansen},
  \citenamefont {Held}, \citenamefont {Adcroft},\ and\ \citenamefont
  {Hallberg}}]{Jansenetal2015}%
  \BibitemOpen
  \bibfield  {author} {\bibinfo {author} {\bibfnamefont {M.~F.}\ \bibnamefont
  {Jansen}}, \bibinfo {author} {\bibfnamefont {I.~M.}\ \bibnamefont {Held}},
  \bibinfo {author} {\bibfnamefont {A.}~\bibnamefont {Adcroft}},\ and\ \bibinfo
  {author} {\bibfnamefont {R.}~\bibnamefont {Hallberg}},\ }\href@noop {}
  {\bibfield  {journal} {\bibinfo  {journal} {Ocean Modell.}\ }\textbf
  {\bibinfo {volume} {94}},\ \bibinfo {pages} {15} (\bibinfo {year}
  {2015})}\BibitemShut {NoStop}%
\bibitem [{\citenamefont {Ding}\ \emph {et~al.}(2026)\citenamefont {Ding},
  \citenamefont {Bobas}, \citenamefont {Scolan}, \citenamefont {Young},\ and\
  \citenamefont {Read}}]{Dingetal2026}%
  \BibitemOpen
  \bibfield  {author} {\bibinfo {author} {\bibfnamefont {S.-S.}\ \bibnamefont
  {Ding}}, \bibinfo {author} {\bibfnamefont {H.}~\bibnamefont {Bobas}},
  \bibinfo {author} {\bibfnamefont {H.}~\bibnamefont {Scolan}}, \bibinfo
  {author} {\bibfnamefont {R.~M.~B.}\ \bibnamefont {Young}},\ and\ \bibinfo
  {author} {\bibfnamefont {P.~L.}\ \bibnamefont {Read}},\ }\href
  {https://doi.org/10.1103/n2nj-dg5r} {\bibfield  {journal} {\bibinfo
  {journal} {Phys. Rev. Lett.}\ }\textbf {\bibinfo {volume} {136}},\ \bibinfo
  {pages} {114101} (\bibinfo {year} {2026})}\BibitemShut {NoStop}%
\bibitem [{\citenamefont {Young}\ and\ \citenamefont
  {Read}(2017)}]{YoungRead2017}%
  \BibitemOpen
  \bibfield  {author} {\bibinfo {author} {\bibfnamefont {R.~M.~B.}\
  \bibnamefont {Young}}\ and\ \bibinfo {author} {\bibfnamefont {P.~L.}\
  \bibnamefont {Read}},\ }\href@noop {} {\bibfield  {journal} {\bibinfo
  {journal} {Nat. Phys.}\ }\textbf {\bibinfo {volume} {13}},\ \bibinfo {pages}
  {1135} (\bibinfo {year} {2017})}\BibitemShut {NoStop}%
\bibitem [{\citenamefont {Thomas}\ \emph {et~al.}(2013)\citenamefont {Thomas},
  \citenamefont {Taylor}, \citenamefont {Ferrari},\ and\ \citenamefont
  {Joyce}}]{Thomasetal2013}%
  \BibitemOpen
  \bibfield  {author} {\bibinfo {author} {\bibfnamefont {L.~N.}\ \bibnamefont
  {Thomas}}, \bibinfo {author} {\bibfnamefont {J.~R.}\ \bibnamefont {Taylor}},
  \bibinfo {author} {\bibfnamefont {R.}~\bibnamefont {Ferrari}},\ and\ \bibinfo
  {author} {\bibfnamefont {T.~M.}\ \bibnamefont {Joyce}},\ }\href@noop {}
  {\bibfield  {journal} {\bibinfo  {journal} {Deep-Sea Res. II}\ }\textbf
  {\bibinfo {volume} {91}},\ \bibinfo {pages} {96} (\bibinfo {year}
  {2013})}\BibitemShut {NoStop}%
\bibitem [{\citenamefont {Chassignet}\ and\ \citenamefont
  {Garraffo}(2001)}]{ChassignetGarraffo2001}%
  \BibitemOpen
  \bibfield  {author} {\bibinfo {author} {\bibfnamefont {E.~P.}\ \bibnamefont
  {Chassignet}}\ and\ \bibinfo {author} {\bibfnamefont {Z.~D.}\ \bibnamefont
  {Garraffo}},\ }in\ \href@noop {} {\emph {\bibinfo {booktitle} {From Stirring
  to Mixing in a Stratified Ocean, Proc. `Aha Huliko`a Hawaiian Winter
  Workshop}}},\ \bibinfo {editor} {edited by\ \bibinfo {editor} {\bibfnamefont
  {P.}~\bibnamefont {M{\"u}ller}}\ and\ \bibinfo {editor} {\bibfnamefont
  {D.}~\bibnamefont {Henderson}}}\ (\bibinfo  {publisher} {University of
  Hawaii},\ \bibinfo {address} {Honolulu},\ \bibinfo {year} {2001})\ pp.\
  \bibinfo {pages} {37--41}\BibitemShut {NoStop}%
\bibitem [{\citenamefont {Galperin}\ and\ \citenamefont
  {Sukoriansky}(2024)}]{GalperinSukoriansky2024}%
  \BibitemOpen
  \bibfield  {author} {\bibinfo {author} {\bibfnamefont {B.}~\bibnamefont
  {Galperin}}\ and\ \bibinfo {author} {\bibfnamefont {S.}~\bibnamefont
  {Sukoriansky}},\ }\href {https://doi.org/10.3390/atmos15080968} {\bibfield
  {journal} {\bibinfo  {journal} {MDPI Atmosphere}\ }\textbf {\bibinfo {volume}
  {15}},\ \bibinfo {eid} {968} (\bibinfo {year} {2024})}\BibitemShut {NoStop}%
\bibitem [{\citenamefont {Pais}(1982)}]{Pais1982}%
  \BibitemOpen
  \bibfield  {author} {\bibinfo {author} {\bibfnamefont {A.}~\bibnamefont
  {Pais}},\ }\href@noop {} {\emph {\bibinfo {title} {Subtle is the Lord: The
  Science and the Life of {A}lbert {E}instein}}}\ (\bibinfo  {publisher}
  {Oxford University Press},\ \bibinfo {address} {Oxford},\ \bibinfo {year}
  {1982})\BibitemShut {NoStop}%
\end{thebibliography}%

\clearpage
\onecolumngrid
\setcounter{table}{0}\setcounter{figure}{0}\setcounter{equation}{0}
\renewcommand{\thetable}{S\arabic{table}}\renewcommand{\thefigure}{S\arabic{figure}}\renewcommand{\theequation}{S\arabic{equation}}
\setlength{\tabcolsep}{4pt}
\begin{center}{\large\bfseries Supplemental Material}\end{center}
\providecommand{\semion}[1]{{\color{blue}[#1]}}
\section*{S1. Data}

\paragraph{SWOT.} We use the SWOT KaRIn Level-2 low-rate sea-surface height product (Expert, version D; NASA PO.DAAC) for August 2023--July 2025, the first two years of the 21-day science orbit (cycles 1--36; cycle 31 is absent and cycle 15 is partial). For each pass we form
\begin{equation}
h=\eta_{\rm SSHA}+\eta_{\rm xover}-\eta_{\rm IT},
\end{equation}
the KaRIn sea-surface height anomaly with the crossover calibration added and the coherent internal tide (HRET model) removed; pixels with a nonzero quality flag are discarded. The field is smoothed over $\pm3$ pixels in both swath directions by normalized convolution (points with less than 80\% valid neighbours are rejected), and geostrophic velocity anomalies are computed from the along- and cross-track gradients, $u_s=-(g/f)\,\partial h/\partial n$, $u_n=(g/f)\,\partial h/\partial s$, then rotated to east/north using the track geometry. Only the swath interior, 12--58~km from nadir, is used.

Velocities are averaged onto a $0.25^\circ$ grid with one sample per pass per bin, so that each pass contributes equally. Eddy second moments $\langle u'u'\rangle$, $\langle v'v'\rangle$, $\langle u'v'\rangle$ are computed per bin about the two-year SWOT mean; bins with fewer than 8 samples are excluded. The numbers of cycles and passes per current are listed in Table~\ref{tab:data}.

\begin{table}[H]\centering\footnotesize
\caption{SWOT coverage per current.}
\label{tab:data}
\begin{adjustbox}{max width=\linewidth}
\begin{tabular}{lccl}\toprule
Current & cycles & passes & box \\ \midrule
Gulf Stream (slope + free jet) & 35 & 2146$^a$ & 25--36$^\circ$N, 82--72$^\circ$W; 33--43$^\circ$N, 77--50$^\circ$W \\
Kuroshio & 35 & 2193 & 26--42$^\circ$N, 128--165$^\circ$E \\
Agulhas & 35 & 1564 & 42--25$^\circ$S, 15--40$^\circ$E \\
East Australian Current & 34 & 860 & 40--24$^\circ$S, 150--162$^\circ$E \\
Brazil Current & 35 & 1719 & 42--20$^\circ$S, 62--35$^\circ$W \\ \bottomrule
\end{tabular}
\end{adjustbox}

\smallskip{\footnotesize $^a$Two overlapping download boxes (slope, 699 pass files; separation and free jet, 1447), merged per cycle.}
\end{table}

\paragraph{GLORYS12.} The mean current is taken from the Copernicus GLORYS12 reanalysis (1/12$^\circ$, daily surface currents), reduced to monthly sums of $u$, $v$, $u^2$, $v^2$, $uv$ and averaged over the SWOT period, August 2023--July 2025 (24 months), for every current. The same monthly sums give the reanalysis' own eddy moments used for comparison.

\section*{S2. Current-following frame and effective viscosity}

\paragraph{Axis.} The axis of each current is traced in the GLORYS mean flow from a seed point upstream: steps of 25~km along the mean velocity, each followed by re-centring on the speed maximum within $\pm120$~km along the local normal; tracing stops where the mean speed falls below 0.05~m\,s$^{-1}$ or at the edge of the box.

\paragraph{Sections and blocks.} At each axis point the unit tangent $\mathbf t$ and the normal $\mathbf n$ (to the left of the flow) define the frame $(s,n)$. The mean along-stream velocity $U_s(n)$ and the eddy stress
\begin{equation}
\langle u_s'u_n'\rangle=-(\langle u'u'\rangle-\langle v'v'\rangle)\sin\theta\cos\theta+\langle u'v'\rangle(\cos^2\theta-\sin^2\theta)
\end{equation}
($\theta$ the direction of $\mathbf t$) are sampled at $n\in[-150,150]$~km every 12.5~km and averaged over 300-km blocks along the path.

\paragraph{Effective viscosity and SNV fit.} On each flank, 25--100~km from the axis,
\begin{equation}
\nu_e=-\langle u_s'u_n'\rangle\big/\langle\partial U_s/\partial n\rangle .
\end{equation}
Across the current ($|n|\le125$~km) we fit
\begin{equation}
\langle u_s'u_n'\rangle=A_2\,\partial_n U_s+A_4\,\partial_n^3U_s
\end{equation}
by least squares. $A_2>0$ is a negative Laplacian viscosity and $A_4>0$ a positive biharmonic one, i.e.\ the SNV structure; $A_2<0$, $A_4<0$ is the opposite (eddy drag).

\paragraph{Flanks.} Flanks are labelled by the sense of the mean shear: the cyclonic flank is to the left of the flow in the Northern Hemisphere and to the right in the Southern Hemisphere. Upstream of separation it is the inshore flank, on the slope.

\paragraph{Uncertainty.} SWOT cycles are resampled with replacement (300 resamplings); each resampling recomputes the eddy moments and all block statistics with the mean flow held fixed. We quote 95\% percentile intervals and, for the fit, the fraction of resamplings with $A_2>0$ and $A_4>0$.

\paragraph{Separation point.} Separation is placed where the current leaves the continental slope: Cape Hatteras (35.9$^\circ$N), the tip of the Agulhas Bank (38.0$^\circ$S), off Boso, Japan (35.1$^\circ$N), the Brazil--Malvinas Confluence (36.2$^\circ$S), and the separation of the East Australian Current (32.9$^\circ$S), where the axis leaves the slope.

\section*{S3. Results for each current}

Tables~\ref{tab:gulf}--\ref{tab:east} list all blocks. Starred values are significant at 95\%. The Florida Straits block of the Gulf Stream (27.2$^\circ$N), where the swath barely spans the channel, is excluded (fit correlation 0.19).

\begin{table}[H]\centering\footnotesize
\caption{Gulf Stream: effective eddy viscosity $\nu_e$ ($10^3$ m$^2$ s$^{-1}$) per 300-km block, with 95\% bootstrap intervals over SWOT cycles; distance $d$ from separation (negative: upstream, on the slope); $P(A_2{>}0)$, $P(A_4{>}0)$: fraction of resamplings; $r$: correlation of the SNV fit.}
\begin{adjustbox}{max width=\linewidth}
\begin{tabular}{rrrr|rr|rrr}\hline
$d$ (km) & lat & lon & $U_{\max}$ & $\nu_e$ cyclonic & $\nu_e$ anticyclonic & $P(A_2{>}0)$ & $P(A_4{>}0)$ & $r$\\\hline
-900 & 29.9 & -79.8 & 1.12 & $+0.4^{*}\,[+0.1,+0.9]$ & $-1.1^{*}\,[-1.6,-0.7]$ & 1.00 & 1.00 & 0.57\\
-600 & 32.0 & -78.2 & 0.85 & $+1.5^{*}\,[+0.2,+3.6]$ & $-1.2^{*}\,[-2.8,-0.0]$ & 0.86 & 1.00 & 0.25\\
-300 & 33.8 & -76.0 & 0.96 & $-3.0^{*}\,[-3.7,-2.1]$ & $-1.3^{*}\,[-1.8,-0.7]$ & 1.00 & 1.00 & 0.56\\
+0 & 35.9 & -74.1 & 1.13 & $+8.6^{*}\,[+6.1,+10.9]$ & $-0.1\,[-1.3,+1.1]$ & 0.00 & 0.00 & 0.71\\
+300 & 37.6 & -71.6 & 0.92 & $+0.0\,[-2.3,+2.4]$ & $+3.0^{*}\,[+0.4,+5.4]$ & 0.01 & 0.11 & 0.69\\
+600 & 37.9 & -68.2 & 0.79 & $-0.9\,[-2.8,+1.1]$ & $-1.9\,[-6.9,+3.3]$ & 0.74 & 0.29 & 0.44\\
+900 & 38.5 & -65.1 & 0.65 & $-0.5\,[-5.5,+4.2]$ & $-4.5\,[-15.5,+5.4]$ & 0.52 & 0.18 & 0.45\\
+1200 & 39.4 & -62.7 & 0.32 & $-31.8\,[-70.9,+2.7]$ & $-5.9\,[-82.8,+76.5]$ & 0.82 & 0.71 & 0.56\\
+1500 & 39.8 & -60.3 & 0.35 & $-12.8\,[-30.9,+4.5]$ & $+13.2\,[-9.4,+34.1]$ & 0.29 & 0.41 & 0.22\\
+1800 & 39.9 & -57.2 & 0.46 & $-6.3\,[-14.8,+3.1]$ & $-0.4\,[-11.8,+10.1]$ & 0.92 & 0.77 & 0.84\\
+2100 & 39.6 & -54.0 & 0.45 & $-9.7^{*}\,[-19.8,-0.1]$ & $+1.9\,[-5.9,+9.6]$ & 0.87 & 0.55 & 0.90\\
+2400 & 39.6 & -51.7 & 0.33 & $-19.7\,[-47.2,+6.6]$ & $+0.8\,[-9.7,+12.1]$ & 0.63 & 0.55 & 0.62\\
\hline\end{tabular}
\end{adjustbox}
\label{tab:gulf}
\end{table}

\begin{table}[H]\centering\footnotesize
\caption{Agulhas: effective eddy viscosity $\nu_e$ ($10^3$ m$^2$ s$^{-1}$) per 300-km block, with 95\% bootstrap intervals over SWOT cycles; distance $d$ from separation (negative: upstream, on the slope); $P(A_2{>}0)$, $P(A_4{>}0)$: fraction of resamplings; $r$: correlation of the SNV fit.}
\begin{adjustbox}{max width=\linewidth}
\begin{tabular}{rrrr|rr|rrr}\hline
$d$ (km) & lat & lon & $U_{\max}$ & $\nu_e$ cyclonic & $\nu_e$ anticyclonic & $P(A_2{>}0)$ & $P(A_4{>}0)$ & $r$\\\hline
-1500 & -30.1 & 31.6 & 0.90 & $-0.3\,[-0.7,+0.2]$ & $+0.7\,[-0.2,+1.8]$ & 0.42 & 0.72 & 0.40\\
-1200 & -32.1 & 29.5 & 1.31 & $-1.0^{*}\,[-1.4,-0.5]$ & $+0.0\,[-0.5,+0.6]$ & 0.93 & 1.00 & 0.86\\
-900 & -33.9 & 27.3 & 1.47 & $+0.0\,[-0.2,+0.3]$ & $+0.2\,[-0.0,+0.5]$ & 0.43 & 0.96 & 0.68\\
-600 & -35.1 & 24.7 & 1.20 & $+1.7^{*}\,[+0.9,+2.5]$ & $+0.7\,[-0.1,+1.8]$ & 0.00 & 0.99 & 0.87\\
-300 & -36.5 & 22.1 & 1.08 & $-0.4\,[-1.1,+0.4]$ & $+1.8^{*}\,[+0.8,+2.8]$ & 0.01 & 0.01 & 0.71\\
+0 & -38.0 & 19.5 & 0.53 & $+11.0^{*}\,[+5.7,+15.9]$ & $+10.1^{*}\,[+2.9,+16.3]$ & 0.00 & 0.24 & 0.95\\
+300 & -37.9 & 16.6 & 0.30 & $-12.7^{*}\,[-21.9,-1.9]$ & $-0.1\,[-12.5,+10.0]$ & 0.91 & 0.24 & 0.87\\
+600 & -39.5 & 16.6 & 0.30 & $-21.0^{*}\,[-40.2,-6.1]$ & $+3.6\,[-3.6,+11.1]$ & 0.19 & 0.01 & -0.02\\
+900 & -40.1 & 19.5 & 0.45 & $-8.7\,[-18.5,+0.9]$ & $+13.8^{*}\,[+5.7,+21.3]$ & 0.00 & 0.07 & 0.93\\
+1200 & -39.6 & 22.6 & 0.58 & $-4.4\,[-10.7,+3.2]$ & $+0.1\,[-7.0,+7.3]$ & 0.68 & 0.41 & 0.82\\
+1500 & -38.1 & 24.5 & 0.48 & $-2.7\,[-15.0,+10.5]$ & $-1.5\,[-9.4,+5.9]$ & 0.64 & 0.62 & 0.37\\
+1800 & -37.5 & 27.2 & 0.44 & $+3.1\,[-5.3,+10.0]$ & $-10.2^{*}\,[-17.8,-2.5]$ & 0.83 & 0.52 & 0.47\\
+2100 & -39.3 & 29.3 & 0.48 & $-6.5^{*}\,[-13.3,-1.2]$ & $+1.0\,[-6.4,+8.4]$ & 0.63 & 0.22 & 0.83\\
+2400 & -39.1 & 31.6 & 0.45 & $-3.6\,[-8.4,+0.9]$ & $-1.0\,[-8.3,+6.9]$ & 0.78 & 0.45 & 0.84\\
+2700 & -38.4 & 34.1 & 0.48 & $+1.4\,[-2.4,+5.4]$ & $+6.0\,[-0.8,+11.5]$ & 0.01 & 0.37 & 0.95\\
+3000 & -39.4 & 36.1 & 0.45 & $-8.0^{*}\,[-11.6,-4.1]$ & $+9.0\,[-1.9,+19.1]$ & 1.00 & 1.00 & 0.55\\
+3300 & -37.9 & 38.0 & 0.51 & $-7.1^{*}\,[-12.0,-2.1]$ & $+0.8\,[-1.3,+3.0]$ & 1.00 & 1.00 & 0.97\\
\hline\end{tabular}
\end{adjustbox}
\label{tab:agulhas}
\end{table}

\begin{table}[H]\centering\footnotesize
\caption{Kuroshio: effective eddy viscosity $\nu_e$ ($10^3$ m$^2$ s$^{-1}$) per 300-km block, with 95\% bootstrap intervals over SWOT cycles; distance $d$ from separation (negative: upstream, on the slope); $P(A_2{>}0)$, $P(A_4{>}0)$: fraction of resamplings; $r$: correlation of the SNV fit.}
\begin{adjustbox}{max width=\linewidth}
\begin{tabular}{rrrr|rr|rrr}\hline
$d$ (km) & lat & lon & $U_{\max}$ & $\nu_e$ cyclonic & $\nu_e$ anticyclonic & $P(A_2{>}0)$ & $P(A_4{>}0)$ & $r$\\\hline
-1800 & 29.6 & 130.5 & 0.24 & $-2.6^{*}\,[-4.6,-0.5]$ & $-3.2\,[-10.6,+5.3]$ & 0.99 & 0.99 & 0.28\\
-1500 & 30.8 & 132.1 & 0.62 & $+0.1\,[-1.4,+1.6]$ & $+3.0\,[-1.7,+8.0]$ & 0.13 & 0.35 & 0.74\\
-1200 & 32.0 & 134.4 & 0.84 & $-1.0\,[-2.6,+0.5]$ & $+5.3^{*}\,[+2.0,+8.7]$ & 0.02 & 0.00 & 0.61\\
-900 & 31.2 & 137.0 & 0.65 & $-2.9\,[-7.3,+1.6]$ & $+2.8\,[-2.8,+8.1]$ & 0.65 & 0.46 & 0.22\\
-600 & 32.6 & 138.5 & 0.58 & $+7.9^{*}\,[+0.0,+16.1]$ & $-5.0^{*}\,[-8.0,-2.0]$ & 0.08 & 0.00 & 0.64\\
-300 & 34.5 & 139.3 & 0.93 & $+1.5^{*}\,[+0.2,+2.9]$ & $-0.6^{*}\,[-1.2,-0.0]$ & 0.03 & 0.05 & 0.25\\
+0 & 35.1 & 141.1 & 0.86 & $+1.2^{*}\,[+0.5,+1.7]$ & $+3.1^{*}\,[+1.4,+4.5]$ & 0.00 & 0.00 & 0.82\\
+300 & 37.3 & 142.5 & 0.56 & $+4.1^{*}\,[+2.7,+5.6]$ & $+12.3^{*}\,[+6.3,+18.3]$ & 0.00 & 0.00 & 0.94\\
+600 & 39.2 & 144.2 & 0.52 & $-2.0\,[-4.7,+0.5]$ & $+5.2\,[-1.4,+11.4]$ & 0.06 & 0.06 & 0.80\\
+900 & 38.0 & 145.7 & 0.56 & $-1.6\,[-7.3,+5.7]$ & $+5.9\,[-0.9,+12.4]$ & 0.05 & 0.34 & 0.92\\
+1200 & 36.3 & 147.0 & 0.81 & $+0.1\,[-2.3,+2.4]$ & $-3.8^{*}\,[-7.6,-0.4]$ & 0.70 & 0.37 & 0.85\\
+1500 & 36.9 & 150.0 & 0.89 & $-1.9\,[-3.7,+0.2]$ & $+1.6\,[-1.4,+4.7]$ & 0.89 & 0.78 & 0.23\\
+1800 & 36.1 & 152.7 & 0.58 & $+1.1\,[-3.6,+6.0]$ & $+5.6\,[-1.8,+13.6]$ & 0.60 & 0.95 & 0.92\\
+2100 & 35.1 & 155.5 & 0.48 & $+7.0\,[-2.7,+16.1]$ & $-2.4\,[-8.2,+2.9]$ & 0.05 & 0.09 & 0.74\\
+2400 & 35.1 & 158.7 & 0.35 & $+8.4\,[-10.6,+27.1]$ & $+2.3\,[-7.2,+13.6]$ & 0.01 & 0.04 & 0.97\\
+2700 & 36.0 & 161.1 & 0.20 & $+34.0\,[-13.9,+78.3]$ & $+1.5\,[-49.3,+45.9]$ & 0.11 & 0.34 & 0.74\\
+3000 & 37.4 & 162.2 & 0.23 & $-1.5\,[-12.2,+9.3]$ & $+33.7\,[-74.4,+137.6]$ & 0.72 & 0.78 & 0.46\\
\hline\end{tabular}
\end{adjustbox}
\label{tab:kuroshio}
\end{table}

\begin{table}[H]\centering\footnotesize
\caption{Brazil Current: effective eddy viscosity $\nu_e$ ($10^3$ m$^2$ s$^{-1}$) per 300-km block, with 95\% bootstrap intervals over SWOT cycles; distance $d$ from separation (negative: upstream, on the slope); $P(A_2{>}0)$, $P(A_4{>}0)$: fraction of resamplings; $r$: correlation of the SNV fit.}
\begin{adjustbox}{max width=\linewidth}
\begin{tabular}{rrrr|rr|rrr}\hline
$d$ (km) & lat & lon & $U_{\max}$ & $\nu_e$ cyclonic & $\nu_e$ anticyclonic & $P(A_2{>}0)$ & $P(A_4{>}0)$ & $r$\\\hline
-2100 & -23.1 & -40.9 & 0.42 & $-1.7^{*}\,[-2.7,-0.6]$ & $+1.3\,[-0.5,+3.3]$ & 0.32 & 0.38 & 0.01\\
-1800 & -23.8 & -43.4 & 0.32 & $-0.5\,[-1.9,+1.1]$ & $+1.3\,[-0.5,+3.1]$ & 0.24 & 0.81 & 0.73\\
-1500 & -25.6 & -45.2 & 0.34 & $-2.2^{*}\,[-3.4,-0.8]$ & $+3.6^{*}\,[+1.6,+5.0]$ & 0.01 & 0.08 & 0.53\\
-1200 & -27.6 & -46.7 & 0.28 & $-2.1^{*}\,[-3.0,-1.1]$ & $+1.0\,[-0.6,+2.9]$ & 0.61 & 0.83 & 0.29\\
-900 & -29.9 & -47.8 & 0.27 & $-0.8\,[-1.7,+0.0]$ & $+0.6\,[-0.6,+1.9]$ & 0.59 & 0.83 & 0.31\\
-600 & -31.9 & -49.3 & 0.29 & $-4.9^{*}\,[-6.2,-3.4]$ & $+2.7^{*}\,[+0.9,+4.3]$ & 0.23 & 0.34 & -0.16\\
-300 & -34.0 & -51.0 & 0.37 & $-1.7^{*}\,[-2.4,-1.1]$ & $+1.8^{*}\,[+0.6,+3.1]$ & 0.45 & 0.34 & 0.13\\
+0 & -36.2 & -52.8 & 0.41 & $+1.7^{*}\,[+0.9,+2.5]$ & $+1.8\,[-0.7,+4.6]$ & 0.05 & 0.49 & 0.65\\
+300 & -38.8 & -53.5 & 0.43 & $+1.3\,[-0.9,+3.6]$ & $+2.1\,[-4.1,+7.7]$ & 0.10 & 0.10 & 0.67\\
+600 & -40.8 & -53.5 & 0.37 & $-15.6^{*}\,[-23.2,-8.0]$ & $+23.7^{*}\,[+13.2,+33.4]$ & 0.00 & 0.06 & 0.77\\
\hline\end{tabular}
\end{adjustbox}
\label{tab:brazil}
\end{table}

\begin{table}[H]\centering\footnotesize
\caption{East Australian Current: effective eddy viscosity $\nu_e$ ($10^3$ m$^2$ s$^{-1}$) per 300-km block, with 95\% bootstrap intervals over SWOT cycles; distance $d$ from separation (negative: upstream, on the slope); $P(A_2{>}0)$, $P(A_4{>}0)$: fraction of resamplings; $r$: correlation of the SNV fit.}
\begin{adjustbox}{max width=\linewidth}
\begin{tabular}{rrrr|rr|rrr}\hline
$d$ (km) & lat & lon & $U_{\max}$ & $\nu_e$ cyclonic & $\nu_e$ anticyclonic & $P(A_2{>}0)$ & $P(A_4{>}0)$ & $r$\\\hline
-600 & -27.8 & 153.9 & 0.63 & $-0.8^{*}\,[-1.2,-0.3]$ & $+3.3^{*}\,[+1.3,+5.3]$ & 0.05 & 0.01 & 0.42\\
-300 & -30.4 & 153.6 & 0.63 & $-0.5\,[-0.9,+0.0]$ & $+4.0^{*}\,[+1.0,+6.9]$ & 0.03 & 0.18 & 0.51\\
+0 & -32.9 & 152.8 & 0.54 & $+11.1^{*}\,[+4.8,+18.0]$ & $+6.8^{*}\,[+1.4,+11.6]$ & 0.00 & 0.00 & 0.90\\
+300 & -35.0 & 151.1 & 0.32 & $-1.4\,[-7.7,+4.7]$ & $+9.8^{*}\,[+2.5,+16.3]$ & 0.00 & 0.00 & 0.77\\
+600 & -36.9 & 150.5 & 0.30 & $-9.4\,[-23.9,+5.8]$ & $+5.6^{*}\,[+0.6,+11.3]$ & 0.09 & 0.15 & 0.69\\
+900 & -38.5 & 150.5 & 0.25 & $-5.9\,[-14.4,+1.6]$ & $+5.5\,[-2.2,+15.8]$ & 0.43 & 0.94 & -0.36\\
\hline\end{tabular}
\end{adjustbox}
\label{tab:east}
\end{table}

\section*{S4. Topographic $\beta$ along the paths}

Depth is taken from the GLORYS12 1/12$^\circ$ model bathymetry (checked against Natural Earth 10-m isobaths). The topographic $\beta$ is $\beta_t=f|\nabla H|/H$, averaged within $\pm50$~km of the axis, and the angle between the current and the $f/H$ contours is computed from the direction of $\nabla(f/H)$. Table~\ref{tab:beta} summarizes the three regimes.

\begin{table}[H]\centering\footnotesize
\caption{Topographic $\beta$ ($\beta_t/\beta_p$) and the angle between current and $f/H$ contours.}
\label{tab:beta}
\begin{adjustbox}{max width=\linewidth}
\begin{tabular}{lccc}\toprule
 & on the slope & at separation & free jet \\ \midrule
$\beta_t/\beta_p$ & 30--115 & falls 10--50$\times$ & 1--30 \\
angle to $f/H$ & 2--11$^\circ$ & 19--37$^\circ$ & 10--70$^\circ$ \\ \bottomrule
\end{tabular}
\end{adjustbox}
\end{table}

\section*{S5. Angular dependence of the negative viscosity (EDQNM)}

To quantify how $\beta$ restricts the inverse transfer to wavevectors aligned with the $\beta$ contours, we computed the two-parametric viscosity $\nu(k|k_c)$ from an eddy-damped quasi-normal Markovian (EDQNM) closure (Orszag 1977, in \emph{Fluid Dynamics}, Les Houches 1973, Gordon and Breach, p.~237) in which the triad relaxation time includes Rossby-wave detuning (Holloway and Hendershott 1977, J.\ Fluid Mech.\ \textbf{82}, 747), ${\rm Re}\,\theta_{kpq}=\mu/(\mu^2+\Delta\omega^2)$, with $\Delta\omega=\omega_k-\omega_p-\omega_q$, $\omega_k=-\beta k_x/k^2$, and eddy damping $\mu_k=\lambda\,(\int_0^k p^2E\,dp)^{1/2}$, $\lambda=0.2$. The transfer is summed over all triads on a $256^2$ lattice with at least one leg beyond $k_c=21$, using an isotropic input spectrum from a $\beta$-plane direct simulation; $\beta$ enters only through the detuning. For $\beta=0$ this gives $\nu=-a[1-1.45(k/k_c)^2]$, close to the SNV form. We evaluate $\nu$ as a function of the angle $\theta$ between the wavevector and the zonal direction. Table~\ref{tab:edqnm} gives $-\nu$ (positive = negative viscosity) in the jet-scale band.

\begin{table}[H]\centering\footnotesize
\caption{EDQNM: $-\nu\times10^4$ in the jet-scale band ($k=1$--4), $\lambda=0.2$, $k_c=21$, versus angle $\theta$ and $\beta$.}
\label{tab:edqnm}
\begin{adjustbox}{max width=\linewidth}
\begin{tabular}{lccccc}\toprule
$\beta$ & 0--10$^\circ$ & 10--20$^\circ$ & 20--30$^\circ$ & 40--50$^\circ$ & 80--90$^\circ$ \\ \midrule
0  & 1.57 & 1.48 & 1.51 & 1.59 & 1.57 \\
3  & 1.57 & 1.39 & 1.20 & 0.82 & 0.54 \\
10 & 1.57 & 0.88 & 0.40 & 0.18 & 0.09 \\
30 & 1.57 & 0.21 & 0.06 & 0.02 & 0.01 \\ \bottomrule
\end{tabular}
\end{adjustbox}
\end{table}

Along the $\beta$ contours the negative viscosity is unaffected; away from them it is switched off, the more sharply the stronger $\beta$. It does not, however, turn positive at jet scales: the closure explains the loss of feeding at separation, while the braking itself comes from the large eddies (S6).

\section*{S6. Scale split of the eddy flux}

Along each SWOT pass the velocity is split by a Gaussian filter of scale $\ell=50$~km along track into large-scale ($u^L$) and small-scale ($u^S=u-u^L$) parts. The stress decomposes exactly as $\langle u'v'\rangle=\langle u^{L\prime}v^{L\prime}\rangle+\langle u^{S\prime}v^{S\prime}\rangle+(\text{cross terms})$, and each part is processed as in S2 (200 resamplings). Tables~\ref{tab:splitgs}--\ref{tab:splitkur} give the cyclonic flank.

Small eddies feed the jet (significantly negative $\nu_e$) in 18 of 30 blocks and are significantly positive in 5. At separation the braking is carried by the large eddies: Gulf Stream at Hatteras, large $+5.6$ vs small $+0.3$; Kuroshio, large $+1.7$ and $+5.4$ vs small $-0.2$ and $-1.1$ ($10^3$~m$^2$\,s$^{-1}$). Caveats: the filter is along track only; the cross terms are not negligible; $\ell=50$~km is a single choice.

\begin{table}[H]\centering\footnotesize
\caption{Gulf Stream: scale-split eddy viscosity on the cyclonic flank ($10^3$ m$^2$ s$^{-1}$), along-track Gaussian split at $\ell=50$ km; in brackets the fraction of 200 resamplings with $\nu_e<0$.}
\begin{adjustbox}{max width=\linewidth}
\begin{tabular}{rrr|rrrr}\hline
$d$ (km) & lat & lon & large & small & cross & total\\\hline
-1200 & 27.2 & -79.7 & $+0.4$ [0.01] & $+0.1$ [0.10] & $+0.2$ [0.02] & $+0.6$ [0.00]\\
-900 & 29.9 & -79.8 & $+0.6$ [0.00] & $-0.2$ [1.00] & $-0.1$ [0.92] & $+0.4$ [0.01]\\
-600 & 32.0 & -78.2 & $+2.9$ [0.00] & $-1.5$ [1.00] & $-0.3$ [0.98] & $+1.0$ [0.14]\\
-300 & 33.8 & -76.0 & $+0.3$ [0.10] & $-3.3$ [1.00] & $-1.0$ [1.00] & $-4.0$ [1.00]\\
+0 & 35.9 & -74.1 & $+5.6$ [0.00] & $+0.3$ [0.22] & $+1.2$ [0.00] & $+7.1$ [0.00]\\
+300 & 37.6 & -71.6 & $+1.3$ [0.08] & $-0.7$ [1.00] & $-0.8$ [0.98] & $-0.2$ [0.55]\\
+600 & 37.9 & -68.2 & $-1.0$ [0.93] & $+0.1$ [0.31] & $+0.1$ [0.38] & $-0.9$ [0.81]\\
+900 & 38.5 & -65.1 & $+1.3$ [0.28] & $-0.6$ [1.00] & $-1.2$ [1.00] & $-0.5$ [0.62]\\
+1200 & 39.4 & -62.7 & $-14.5$ [0.85] & $-4.2$ [1.00] & $-13.5$ [1.00] & $-32.2$ [0.96]\\
+1500 & 39.8 & -60.3 & $-9.3$ [0.90] & $-1.4$ [1.00] & $-3.0$ [0.93] & $-13.7$ [0.93]\\
+1800 & 39.9 & -57.2 & $-2.5$ [0.76] & $-1.4$ [1.00] & $-3.2$ [1.00] & $-7.1$ [0.91]\\
+2100 & 39.6 & -54.0 & $-9.7$ [0.98] & $+0.4$ [0.03] & $+0.1$ [0.46] & $-9.2$ [0.97]\\
+2400 & 39.6 & -51.7 & $-14.4$ [0.85] & $-2.5$ [1.00] & $-4.6$ [0.93] & $-21.4$ [0.91]\\
\hline\end{tabular}
\end{adjustbox}
\label{tab:splitgs}\end{table}

\begin{table}[H]\centering\footnotesize
\caption{Kuroshio: scale-split eddy viscosity on the cyclonic flank ($10^3$ m$^2$ s$^{-1}$), along-track Gaussian split at $\ell=50$ km; in brackets the fraction of 200 resamplings with $\nu_e<0$.}
\begin{adjustbox}{max width=\linewidth}
\begin{tabular}{rrr|rrrr}\hline
$d$ (km) & lat & lon & large & small & cross & total\\\hline
-1800 & 29.6 & 130.5 & $-1.8$ [0.99] & $+0.6$ [0.10] & $-1.8$ [1.00] & $-3.0$ [0.98]\\
-1500 & 30.8 & 132.1 & $+3.0$ [0.00] & $-2.7$ [1.00] & $-1.0$ [1.00] & $-0.6$ [0.73]\\
-1200 & 32.0 & 134.4 & $-0.1$ [0.60] & $-0.4$ [1.00] & $-0.6$ [1.00] & $-1.1$ [0.90]\\
-900 & 31.2 & 137.0 & $-3.7$ [0.96] & $+0.4$ [0.01] & $+0.5$ [0.17] & $-2.8$ [0.88]\\
-600 & 32.6 & 138.5 & $+8.3$ [0.01] & $-0.7$ [1.00] & $+0.1$ [0.39] & $+7.7$ [0.04]\\
-300 & 34.5 & 139.3 & $-1.0$ [0.94] & $+2.7$ [0.00] & $+1.1$ [0.00] & $+2.8$ [0.00]\\
+0 & 35.1 & 141.1 & $+1.7$ [0.00] & $-0.2$ [0.95] & $+0.3$ [0.00] & $+1.7$ [0.00]\\
+300 & 37.3 & 142.5 & $+5.4$ [0.00] & $-1.1$ [1.00] & $-0.4$ [0.99] & $+3.9$ [0.00]\\
+600 & 39.2 & 144.2 & $-0.5$ [0.63] & $-1.0$ [1.00] & $-0.8$ [1.00] & $-2.3$ [0.98]\\
+900 & 38.0 & 145.7 & $-1.6$ [0.72] & $-0.1$ [0.69] & $-0.1$ [0.67] & $-1.7$ [0.72]\\
+1200 & 36.3 & 147.0 & $+0.2$ [0.38] & $-0.1$ [0.80] & $-0.1$ [0.62] & $-0.0$ [0.46]\\
+1500 & 36.9 & 150.0 & $-0.9$ [0.90] & $-0.4$ [1.00] & $-0.7$ [1.00] & $-2.1$ [0.99]\\
+1800 & 36.1 & 152.7 & $-1.6$ [0.72] & $+1.1$ [0.00] & $+1.6$ [0.00] & $+1.0$ [0.32]\\
+2100 & 35.1 & 155.5 & $+4.3$ [0.14] & $+0.9$ [0.00] & $+2.1$ [0.02] & $+7.3$ [0.07]\\
+2400 & 35.1 & 158.7 & $+7.9$ [0.18] & $-0.2$ [0.68] & $+0.8$ [0.32] & $+8.5$ [0.17]\\
+2700 & 36.0 & 161.1 & $+37.0$ [0.04] & $-2.9$ [1.00] & $-0.6$ [0.53] & $+33.5$ [0.10]\\
+3000 & 37.4 & 162.2 & $+1.5$ [0.38] & $-1.4$ [1.00] & $-2.3$ [0.99] & $-2.2$ [0.68]\\
\hline\end{tabular}
\end{adjustbox}
\label{tab:splitkur}\end{table}

\section*{S7. Spectral test of the SNV form}
We use GLORYS12 daily surface currents (Gulf Stream 2015--2016; Kuroshio Extension and ACC 2015) in three boxes: Gulf Stream (34--42$^\circ$N, 72--48$^\circ$W), Kuroshio Extension (32--40$^\circ$N, 145--170$^\circ$E) and Antarctic Circumpolar Current south of Kerguelen (58--44$^\circ$S, 75--125$^\circ$E). Velocities are Tukey-windowed, and for each month the subgrid transfer to resolved wavenumbers $k<k_c$ is computed from the two-dimensional vorticity triads and expressed as $\nu(k|k_c)=-T(k|k_c)/(2k^2E(k))$, where $T(k|k_c)$ is the energy transfer; this is equivalent to the enstrophy-based definition of Chekhlov \emph{et al.}\ (1996, Eq.~9). The fit $\nu(k|k_c)=-a[1-c\,(k/k_c)^2]$ is made for cutoffs $L_c=2\pi/k_c=100$--300~km. The transfer is averaged over wavenumber shells, i.e. over all directions of the wavevector, so the fit tests the isotropic SNV form. Chekhlov \emph{et al.} (1996) resolved $\nu(k,\varphi|k_c)$ in direction in $\beta$-plane simulations: it stays negative near the zonal wavevectors ($k_x=0$) and vanishes elsewhere. With a few shells per fit in the present boxes the angular dependence cannot be resolved; it is a natural test for higher-resolution data over larger domains.

The negative Laplacian part is present everywhere: $a>0$ in all 48 months in all three basins. The shape parameter $c$ depends on the cutoff, but collapses when the cutoff is scaled with the deformation radius $L_d$ (computed from the measured $N^2$ profiles through the first baroclinic mode: Gulf Stream 27.3~km, Kuroshio 30.6~km, ACC 14.3~km):
\begin{equation}
c=3.72-1.03\ln(L_c/L_d),\qquad R^2=0.94,
\end{equation}
with $c=8/5$ at $L_c\approx8L_d$. The amplitude $a$ exceeds the isotropic estimate $(25/18)\,\Pi_\varepsilon/\bar\Omega$ by a factor of 4.6--16 and follows the energy flux rather than $1/\bar\Omega$: in the Gulf Stream, with $\Pi_\varepsilon$ the inverse energy flux at 250--400~km computed each month from the same transfer ($2.0$--$4.1\times10^{-8}$~W\,kg$^{-1}$), $a$ correlates with $\Pi_\varepsilon$ ($r=0.61$) but not with $1/\bar\Omega$ ($r=-0.5$), consistent with energy injection near $L_d$ rather than at small scales. Averaged over each box, $\Pi_\varepsilon$ is inverse in every month; its local sign along the current paths was not resolved by this box-averaged test. Part of the excess may also reflect the direction averaging, which mixes zonal and non-zonal wavevectors.

\section*{S8. The SNV closure}

We summarize the derivation of Eq.~(1) of the Letter, following Sukoriansky \emph{et al.}\ (1996, J.\ Sci.\ Comput.\ \textbf{11}, 13; hereafter S96) for isotropic two-dimensional turbulence without $\beta$.

\paragraph{Two-parametric viscosity.} As elaborated in S96, the SNV scheme is derived using the 2D forced dissipative
vorticity equation,
\begin{equation}
{\partial \zeta \over \partial t} + {\partial \left (\nabla^{-2} \zeta, 
\zeta \right ) \over \partial (x, y) } = \nu_o \nabla^2 \zeta + \xi. \label{one} 
\end{equation}
Here, $\zeta$ is the fluid vorticity, $\nu_o$ is the molecular viscosity, and $\xi$ is the forcing; $x$ and $y$
are coordinate axes. We set the cutoff wavenumber at $k_c$ such that only the modes $k<k_c$ are 
resolved. Their interaction with the subgrid modes $k>k_c$ is described by the two-parametric viscosity introduced by Kraichnan (1976, J.\ Atmos.\ Sci.\ \textbf{33}, 1521; hereafter K76),
\begin{equation}
\nu(k|k_c)=-\frac{T_\Omega(k|k_c)}{2k^2\Omega(k)},
\end{equation}
where $\Omega(k)=k^2E(k)$ is the enstrophy spectrum and $T_\Omega(k|k_c)$ is the enstrophy transfer rate
into the mode $k$ by all triads with at least one leg residing beyond $k_c$. No scale separation is assumed
such that $\nu$ depends on both $k$ and $k_c$. When energy is injected on unresolved scales beyond 
$k_c$, $\nu(k|k_c)$ is negative and nearly constant for $k\ll k_c$ as resolved modes receive energy by virtue of the inverse cascade, and positive near $k_c$, where they lose enstrophy to the subgrid scales. This shape was obtained from the test-field model (K76) and from renormalization-group theory, and confirmed in direct numerical simulation by Chekhlov \emph{et al.}\ (1994, Phys.\ Fluids \textbf{6}, 2548).

\paragraph{The energy feedback.} If the energy spectrum in the resolved subrange is $E(k,t)$, the energy
received by the modes in this subrange is
\begin{equation}
\widehat{\Pi}_\varepsilon(t) = -2\int_0^{k_c}\nu(k|k_c)\,k^2E(k,t)\,dk,   \label{eq:eLES}
\end{equation}
positive when energy is transferred to the resolved scales (the opposite of the usual sign convention for the spectral flux). The positive feedback between $\widehat{\Pi}_\varepsilon(t)$ and the resolved enstrophy 
$\bar\Omega(t)=\int_0^{k_c}k^2E\,dk$ renders this closure unstable (S96, Sec.~4.2). It is stabilized 
by setting $\nu(k|k_c) = F(\bar\Omega)\,N(k/k_c)$ and choosing $F$ such that $\widehat{\Pi}_\varepsilon(t)$
is equal to the energy transfer across $k_c$, $\Pi_\varepsilon(t)$, at all times; in the steady state of S96 this is the rate of energy injection. Then, the inverse
proportionality between $F$ and $\bar\Omega$ turns the feedback negative.

\paragraph{In the ocean.} S96 simulated a statistically steady state, in which $\widehat{\Pi}_\varepsilon$
could be specified. In the ocean there is no steady state, and the energy input to the resolved scales 
cannot be imposed: it is the instantaneous flux $\Pi_\varepsilon(k_c,t)$ across the cutoff, delivered by the flow, which varies in time and space and can change sign. The derivation below then holds instant by instant, as a diagnostic relation rather than a constraint. Where the transfer across $k_c$ is inverse, $\Pi_\varepsilon>0$
and the Laplacian viscosity is negative; where it reverses, $\Pi_\varepsilon<0$, the Laplacian viscosity becomes positive and the eddies brake the flow. This is our interpretation of the braking where the jets separate from the coast (S7: the local sign of $\Pi_\varepsilon$ along the paths is not yet measured).

\paragraph{Two-term form.} Expanding the normalized profile in powers of $k^2$ and keeping two terms, 
one obtains
\begin{equation}
N(k/k_c)\simeq-1+c\,(k/k_c)^2.
\end{equation}
For a Kolmogorov spectrum $E(k) = C_K \Pi_\varepsilon^{2/3} k^{-5/3}$, two conditions fix $c$ and $F$ for
the resolved modes:
\begin{itemize}
\item[(i)] zero enstrophy transfer in the energy subrange, $\int_0^{k_c}\nu(k|k_c)\,k^4E\,dk=0$. With $\int_0^{k_c}k^4E\,dk\propto\tfrac{3}{10}k_c^{10/3}$ and $\int_0^{k_c}k^6E\,dk\propto\tfrac{3}{16}k_c^{16/3}$, this gives $c=8/5$;
\item[(ii)] the energy input equals the transfer across the cutoff, $\widehat{\Pi}_\varepsilon = \Pi_\varepsilon$.  Along with 
$\int_0^{k_c}k^4E\,dk=\tfrac25\,\bar\Omega\,k_c^2$, Eq.~(\ref{eq:eLES}) gives 
$\Pi_\varepsilon=2F\bar\Omega\,(1-\tfrac{8}{5}\cdot\tfrac{2}{5})=\tfrac{18}{25}F\bar\Omega$, so $F=25\Pi_\varepsilon/(18\bar\Omega)$.
\end{itemize}
Together, (i) and (ii) yield
\begin{equation}
\nu(k|k_c)=-a\left[1-c\left(\frac{k}{k_c}\right)^2\right],\qquad a=\frac{25}{18}
\frac{\Pi_\varepsilon}{\bar\Omega},\qquad c=\frac85 ,
\end{equation}
which is Eq.~(1) of the Letter. LES with this two-term form reproduce the Kolmogorov spectrum and the conservation laws as well as LES with the full $N(k/k_c)$ (S96, Sec.~5).

\paragraph{Physical space.} Since $\nu k^2\hat\zeta$ corresponds to $-\nu\nabla^2\zeta$, the subgrid term becomes
\begin{equation}
{\partial \zeta \over \partial t} + {\partial \left (\nabla^{-2} \zeta, \zeta \right ) \over \partial (x, y) }=-\nabla\!\cdot\!(A_2\nabla\zeta)-A_4\nabla^4\zeta ,
\qquad A_2=\frac{25}{18}\frac{\Pi_\varepsilon}{\bar\Omega},\qquad A_4=\frac{c\,A_2}{k_c^2}=\frac{20}{9}\frac{\Pi_\varepsilon}{\bar\Omega k_c^2}.
\end{equation}
Both coefficients are positive: a negative Laplacian viscosity, which feeds the resolved scales, and a positive biharmonic viscosity, which removes enstrophy near the cutoff. In S96 $\bar\Omega$ is the enstrophy averaged over a grid cell, and the Laplacian term is written in conservative form. Because the biharmonic term acts mainly near $k_c$, it can be held constant by inserting the Kolmogorov value 
$\bar\Omega=\tfrac34 C_K \Pi_\varepsilon^{2/3}k_c^{4/3}$, which gives $A_4=0.511\, \Pi_\varepsilon^{1/3}
k_c^{-10/3}$ for $C_K\simeq5.8$; the negative Laplacian term must remain flow-dependent (S96, Sec.~5.2).

For a mean flow $U(n)$ the same operator gives $\langle u'v'\rangle=A_2\,\partial_nU+A_4\,\partial_n^3U$, the form fitted in S2, with $A_2>0$ and $A_4>0$ for SNV. The linear growth rate $A_2k^2-A_4k^4$ is largest at $k_*=k_c/\sqrt{2c}=k_c/\sqrt{3.2}$, the fastest-growing wavenumber of the Kuramoto--Sivashinsky operator quoted in the Letter.

\paragraph{Assumptions and limits.} The derivation assumes isotropy, energy injection well inside the subgrid range ($k_c\ll k_f$), and a Kolmogorov spectrum of the resolved modes. In the ocean the energy is injected near the deformation radius, so the second and third assumptions hold only for $L_c\gg L_d$; this is why $c$ approaches $8/5$ only at large $L_c/L_d$ (S7). The closure describes the resolved flow relative to the cutoff: if all scales are resolved and the forcing is explicit, both eddy terms vanish and only the positive molecular viscosity remains (S96, Sec.~6). On the $\beta$-plane the two-parametric viscosity becomes anisotropic (Chekhlov \emph{et al.} 1996). It is derived in the same way as in the isotropic case, the only difference being that the triad relaxation time depends on $\beta$ (Huang \emph{et al.} 2001, Phys.\ Fluids \textbf{13}, 225, Eqs.~2.7--2.10); its angular dependence is discussed in S5.

\section*{S9. Kuramoto--Sivashinsky equation}

The one-dimensional Kuramoto--Sivashinsky equation, $u_t+uu_x+u_{xx}+u_{xxxx}=0$, has the SNV operator as its fixed linear part. We integrated it (ETDRK4, $L=400$, $N=2048$, $T=600$) and computed the a-priori subgrid eddy viscosity $\nu(k|k_c)$ from the resolved and subgrid modes. At large scales it is mostly positive ($+0.2$ to $+1.1$, against the bare $-1$), in agreement with Yakhot (1981): in one dimension there is no inverse cascade, and the chaotic large scales behave as a noisy Burgers flow with positive viscosity. Negative eddy viscosity is therefore a property of two-dimensional turbulence, not of the operator's form.

\section*{S10. Potential-vorticity mixing test}

Down-gradient mixing of potential vorticity with diffusivity $\kappa$, $\langle v'q'\rangle=-\kappa\,\partial_n Q$, with $\partial_nQ\simeq\beta_t-\partial_n^2U_s$ on the slope, implies through the Taylor identity $\langle u_s'u_n'\rangle=\kappa\beta_t\,n-\kappa\,\partial_nU_s+\text{const}$: a term linear in $n$ and a term $-\kappa\,\partial_nU_s$ that the fit reads as a positive viscosity. To test whether the SNV signs reflect this, each block is refitted as
\begin{equation}
\langle u_s'u_n'\rangle=A_0+A_1n+A_2\,\partial_nU_s+A_4\,\partial_n^3U_s ,
\end{equation}
with the regression columns normalized, and the SWOT cycles are resampled 200 times as in S2. Tables~\ref{t:first}--\ref{t:last} give, for every block, the fraction of resamplings with $A_2>0$ and $A_4>0$ without and with the added terms.

\textbf{Summary.} The SNV structure survives the PV-mixing terms in nearly all resamplings at 29.9$^\circ$N in the Gulf Stream, at the Loop Current's reattachment to the West Florida slope and in the Florida Straits, and in the Agulhas Return Current ($+3000$ and $+3300$~km); in most resamplings on the Agulhas slope at 32.1$^\circ$S (0.88/0.77) and in the Kuroshio at 34.5$^\circ$N (0.94/0.94). It does not survive at 33.8$^\circ$N in the Gulf Stream or at the Kuroshio's exit from the East China Sea (29.6$^\circ$N), where the linear term and a positive viscosity account for the flux. In the Brazil and East Australian Currents the SNV structure is not resolved in either fit. Braking at separation ($A_2<0$) survives in the Gulf Stream, Kuroshio and Agulhas, and is ambiguous in the Brazil and East Australian Currents. In GLORYS's own fluxes the SNV signs survive in all four Gulf Stream slope blocks; the SWOT results are those quoted in the main text. The scale split (S6) is an independent discriminator: down-gradient PV mixing does not predict that the up-gradient flux is carried by the small eddies.

\begin{table}[H]\centering\footnotesize
\caption{Gulf Stream, SWOT: fraction of 200 resamplings with $A_2>0$ and $A_4>0$, without and with the PV-mixing terms; $d$ from separation.}
\begin{adjustbox}{max width=\linewidth}
\begin{tabular}{rr|cc|cc}\toprule
$d$ (km) & lat & \multicolumn{2}{c|}{SNV only} & \multicolumn{2}{c}{with $A_0+A_1n$}\\
 & & $P(A_2{>}0)$ & $P(A_4{>}0)$ & $P(A_2{>}0)$ & $P(A_4{>}0)$\\ \midrule
$-1200$ & 27.2 & 0.00 & 0.00 & 1.00 & 1.00\\
$-900$ & 29.9 & 1.00 & 1.00 & 1.00 & 1.00\\
$-600$ & 32.0 & 0.89 & 0.99 & 0.48 & 0.83\\
$-300$ & 33.8 & 1.00 & 1.00 & 0.00 & 0.00\\
$+0$ & 35.9 & 0.00 & 0.00 & 0.00 & 0.00\\
$+300$ & 37.6 & 0.02 & 0.10 & 0.09 & 0.22\\
$+600$ & 37.9 & 0.77 & 0.28 & 0.98 & 0.97\\
$+900$ & 38.5 & 0.54 & 0.21 & 0.79 & 0.69\\
$+1200$ & 39.4 & 0.81 & 0.73 & 0.47 & 0.47\\
$+1500$ & 39.8 & 0.21 & 0.45 & 0.01 & 0.01\\
$+1800$ & 39.9 & 0.90 & 0.72 & 0.42 & 0.26\\
$+2100$ & 39.6 & 0.84 & 0.53 & 0.61 & 0.56\\
$+2400$ & 39.6 & 0.64 & 0.64 & 0.55 & 0.80\\
\bottomrule\end{tabular}
\end{adjustbox}\label{t:first}\end{table}
\begin{table}[H]\centering\footnotesize
\caption{Kuroshio, SWOT: fraction of 200 resamplings with $A_2>0$ and $A_4>0$, without and with the PV-mixing terms; $d$ from separation.}
\begin{adjustbox}{max width=\linewidth}
\begin{tabular}{rr|cc|cc}\toprule
$d$ (km) & lat & \multicolumn{2}{c|}{SNV only} & \multicolumn{2}{c}{with $A_0+A_1n$}\\
 & & $P(A_2{>}0)$ & $P(A_4{>}0)$ & $P(A_2{>}0)$ & $P(A_4{>}0)$\\ \midrule
$-1800$ & 29.6 & 1.00 & 1.00 & 0.00 & 0.00\\
$-1500$ & 30.8 & 0.17 & 0.39 & 0.47 & 0.66\\
$-1200$ & 32.0 & 0.04 & 0.01 & 0.74 & 0.47\\
$-900$ & 31.2 & 0.70 & 0.51 & 0.46 & 0.17\\
$-600$ & 32.6 & 0.10 & 0.01 & 0.65 & 0.30\\
$-300$ & 34.5 & 0.04 & 0.07 & 0.94 & 0.94\\
$+0$ & 35.1 & 0.00 & 0.00 & 0.07 & 0.00\\
$+300$ & 37.3 & 0.00 & 0.00 & 0.00 & 0.00\\
$+600$ & 39.2 & 0.05 & 0.07 & 0.39 & 0.29\\
$+900$ & 38.0 & 0.06 & 0.36 & 0.25 & 0.59\\
$+1200$ & 36.3 & 0.75 & 0.33 & 0.68 & 0.26\\
$+1500$ & 36.9 & 0.84 & 0.74 & 0.67 & 0.64\\
$+1800$ & 36.1 & 0.61 & 0.96 & 0.07 & 0.73\\
$+2100$ & 35.1 & 0.06 & 0.10 & 0.31 & 0.32\\
$+2400$ & 35.1 & 0.04 & 0.06 & 0.30 & 0.22\\
$+2700$ & 36.0 & 0.12 & 0.37 & 0.06 & 0.07\\
$+3000$ & 37.4 & 0.74 & 0.83 & 0.65 & 0.80\\
\bottomrule\end{tabular}
\end{adjustbox}\end{table}
\begin{table}[H]\centering\footnotesize
\caption{Agulhas, SWOT: fraction of 200 resamplings with $A_2>0$ and $A_4>0$, without and with the PV-mixing terms; $d$ from separation.}
\begin{adjustbox}{max width=\linewidth}
\begin{tabular}{rr|cc|cc}\toprule
$d$ (km) & lat & \multicolumn{2}{c|}{SNV only} & \multicolumn{2}{c}{with $A_0+A_1n$}\\
 & & $P(A_2{>}0)$ & $P(A_4{>}0)$ & $P(A_2{>}0)$ & $P(A_4{>}0)$\\ \midrule
$-1500$ & -30.1 & 0.47 & 0.72 & 0.14 & 0.54\\
$-1200$ & -32.1 & 0.94 & 1.00 & 0.88 & 0.77\\
$-900$ & -33.9 & 0.47 & 0.95 & 0.03 & 0.59\\
$-600$ & -35.1 & 0.00 & 0.98 & 0.00 & 0.13\\
$-300$ & -36.5 & 0.01 & 0.01 & 0.03 & 0.01\\
$+0$ & -38.0 & 0.00 & 0.26 & 0.00 & 0.01\\
$+300$ & -37.9 & 0.93 & 0.24 & 0.41 & 0.28\\
$+600$ & -39.5 & 0.18 & 0.01 & 0.90 & 0.73\\
$+900$ & -40.1 & 0.00 & 0.04 & 0.20 & 0.45\\
$+1200$ & -39.6 & 0.78 & 0.46 & 0.75 & 0.74\\
$+1500$ & -38.1 & 0.66 & 0.56 & 0.76 & 0.82\\
$+1800$ & -37.5 & 0.76 & 0.55 & 0.89 & 0.70\\
$+2100$ & -39.3 & 0.59 & 0.27 & 0.91 & 0.86\\
$+2400$ & -39.1 & 0.81 & 0.40 & 0.82 & 0.69\\
$+2700$ & -38.4 & 0.02 & 0.33 & 0.24 & 0.36\\
$+3000$ & -39.4 & 1.00 & 1.00 & 0.99 & 1.00\\
$+3300$ & -37.9 & 1.00 & 1.00 & 0.91 & 0.96\\
\bottomrule\end{tabular}
\end{adjustbox}\end{table}
\begin{table}[H]\centering\footnotesize
\caption{Brazil Current, SWOT: fraction of 200 resamplings with $A_2>0$ and $A_4>0$, without and with the PV-mixing terms; $d$ from separation.}
\begin{adjustbox}{max width=\linewidth}
\begin{tabular}{rr|cc|cc}\toprule
$d$ (km) & lat & \multicolumn{2}{c|}{SNV only} & \multicolumn{2}{c}{with $A_0+A_1n$}\\
 & & $P(A_2{>}0)$ & $P(A_4{>}0)$ & $P(A_2{>}0)$ & $P(A_4{>}0)$\\ \midrule
$-2100$ & -23.1 & 0.22 & 0.25 & 0.81 & 0.85\\
$-1800$ & -23.8 & 0.18 & 0.77 & 0.00 & 0.23\\
$-1500$ & -25.6 & 0.01 & 0.09 & 0.01 & 0.03\\
$-1200$ & -27.6 & 0.64 & 0.82 & 0.26 & 0.12\\
$-900$ & -29.9 & 0.55 & 0.79 & 0.04 & 0.08\\
$-600$ & -31.9 & 0.20 & 0.34 & 0.36 & 0.41\\
$-300$ & -34.0 & 0.41 & 0.33 & 0.18 & 0.10\\
$+0$ & -36.2 & 0.05 & 0.47 & 0.07 & 0.70\\
$+300$ & -38.8 & 0.10 & 0.10 & 0.04 & 0.00\\
$+600$ & -40.8 & 0.01 & 0.04 & 0.24 & 0.00\\
\bottomrule\end{tabular}
\end{adjustbox}\end{table}
\begin{table}[H]\centering\footnotesize
\caption{East Australian Current, SWOT: fraction of 200 resamplings with $A_2>0$ and $A_4>0$, without and with the PV-mixing terms; $d$ from separation.}
\begin{adjustbox}{max width=\linewidth}
\begin{tabular}{rr|cc|cc}\toprule
$d$ (km) & lat & \multicolumn{2}{c|}{SNV only} & \multicolumn{2}{c}{with $A_0+A_1n$}\\
 & & $P(A_2{>}0)$ & $P(A_4{>}0)$ & $P(A_2{>}0)$ & $P(A_4{>}0)$\\ \midrule
$-600$ & -27.8 & 0.01 & 0.01 & 0.00 & 0.00\\
$-300$ & -30.4 & 0.02 & 0.17 & 0.05 & 0.10\\
$+0$ & -32.9 & 0.00 & 0.00 & 0.32 & 0.65\\
$+300$ & -35.0 & 0.00 & 0.00 & 0.89 & 0.85\\
$+600$ & -36.9 & 0.04 & 0.10 & 0.03 & 0.04\\
$+900$ & -38.5 & 0.36 & 0.96 & 0.64 & 0.42\\
\bottomrule\end{tabular}
\end{adjustbox}\end{table}
\begin{table}[H]\centering\footnotesize
\caption{Loop Current, SWOT (35 cycles; matched 24-month mean): fraction of 200 resamplings with $A_2>0$ and $A_4>0$.}
\begin{adjustbox}{max width=\linewidth}
\begin{tabular}{l|cc|cc}\toprule
block & \multicolumn{2}{c|}{SNV only} & \multicolumn{2}{c}{with $A_0+A_1n$}\\
 & $P(A_2{>}0)$ & $P(A_4{>}0)$ & $P(A_2{>}0)$ & $P(A_4{>}0)$\\ \midrule
Yucatan & 0.01 & 0.40 & 0.00 & 0.07\\
leaves Campeche Bank & 0.00 & 0.00 & 0.47 & 0.78\\
Loop tip & 0.99 & 0.98 & 0.27 & 0.23\\
east limb & 0.28 & 0.02 & 0.28 & 0.09\\
Dry Tortugas (West Florida slope) & 1.00 & 1.00 & 1.00 & 1.00\\
Florida Straits & 1.00 & 0.93 & 1.00 & 1.00\\ \bottomrule
\end{tabular}
\end{adjustbox}
\label{t:last}\end{table}

\section*{Code and data availability}

The scripts used for the SWOT and GLORYS analyses and the derived block statistics will be deposited in a public repository upon acceptance; until then they are available from the authors on request.

\end{document}